\documentclass[aps,floats]{revtex4}
\usepackage{amsmath,amssymb}
\usepackage{graphicx,epsfig}
\usepackage[greek,english]{babel}
\usepackage{bbold}

\begin{document}
\bibliographystyle {plain}
\pdfoutput=1
\def\oppropto{\mathop{\propto}} 
\def\opsimeq{\mathop{\simeq}}
\def\opoverderline{\mathop{\overline}}
\def\operarrow{\mathop{\longrightarrow}}
\def\opsim{\mathop{\sim}}

\def\opmin{\mathop{\min}} 
\def\opmax{\mathop{\max}} 
\def\oplim{\mathop{\lim}}

\def\fig#1#2{\includegraphics[height=#1]{#2}}
\def\figx#1#2{\includegraphics[width=#1]{#2}}


\title{ Large deviations at various levels for run-and-tumble processes   \\
with space-dependent velocities and space-dependent switching rates } 


\author{ C\'ecile Monthus }
 \affiliation{Institut de Physique Th\'{e}orique, 
Universit\'e Paris Saclay, CNRS, CEA,
91191 Gif-sur-Yvette, France}

\begin{abstract}

One-dimensional run-and-tumble processes may converge towards some localized non-equilibrium steady state when the two velocities and/or the two switching rates are space-dependent. A long dynamical trajectory can be then analyzed via the large deviations at Level 2.5 for the joint probability of the empirical densities, of the empirical spatial currents and of the empirical switching flows. The Level 2 for the empirical densities alone can be then derived via the optimization of the Level 2.5 over the empirical flows. More generally, the large deviations of any time-additive observable can be also obtained via contraction from the Level 2.5, or equivalently via the deformed generator method and the corresponding Doob conditioned process. Finally,  the large deviations for the empirical intervals between consecutive switching events can be obtained via the introduction of the alternate Markov chain that governs the series of all the switching events of a long trajectory.

\end{abstract}

\maketitle


\section{ Introduction}

Among the various stochastic intermittent dynamics that have attracted a lot of interest   
recently (see the reviews \cite{review_search,review_reset} and references therein),
run-and-tumble processes have played a major role 
and a lot of their properties have been studied in many different situations 
\cite{old,tailleur2008,angelani2014,angelani2015,evans2016,evans2017,redner2018,satya2018,evans2018,emil2018,pld2019,harris,emil2019,kundu2019,angelani,emil2020,satya2019,hartmann2020,satya2020,rosso2020,gregory2019,kundu2020,pld2020,entropyprod,bressloff2020,localtime,gregory2021a,boyer2021,bressloff2021,gregory2021b}.
In the present paper, the goal is to analyze the large deviations properties of the one-dimensional run-and-tumble process on the infinite line, when the space-dependence of the two velocities and/or of the two switching rates
 produces a localized non-equilibrium steady state.

Within the theory of large deviations  
(see the reviews \cite{oono,ellis,review_touchette} and references therein),
the traditional classification into three nested levels for dynamical trajectories 
over a large time-window $T$ \cite{oono,review_touchette}, 
with the Level 1 for empirical time-averaged observables of the position, 
the Level 2 for the empirical time-averaged density of the position,
and the Level 3 for the whole empirical process,
has turned out to be inadequate for non-equilibrium steady states where steady currents play a major role,
because the Level 2 is insufficient and not closed, while the Level 3 is too general.
The introduction of the Level 2.5 concerning the joint probability distribution of the empirical time-averaged density 
of the position and of the empirical time-averaged flows has thus been a major achievement.
 Indeed, the rate functions at Level 2.5 are explicit for various types of Markov processes,
including discrete-time Markov Chains
 \cite{fortelle_thesis,fortelle_chain,review_touchette,c_largedevdisorder,c_reset,c_inference},
continuous-time Markov Jump processes
\cite{fortelle_thesis,fortelle_jump,maes_canonical,maes_onandbeyond,wynants_thesis,chetrite_formal,BFG1,BFG2,chetrite_HDR,c_ring,c_interactions,c_open,c_detailed,barato_periodic,chetrite_periodic,c_reset,c_inference}
and Diffusion processes described by Fokker-Planck equations
\cite{wynants_thesis,maes_diffusion,chetrite_formal,engel,chetrite_HDR,c_reset,c_lyapunov,c_inference}.
As a consequence, the explicit Level 2.5 plays a central role for any Markov process
converging towards some non-equilibrium steady state, and many other large deviations properties can be derived
from the Level 2.5 via contraction procedures.
The first important example is of course the Level 2 for the empirical density alone 
that should be obtained via the optimization of the Level 2.5 over the empirical flows.
More generally, the Level 2.5 can be contracted
to obtain the large deviations properties of any time-additive observable
of the dynamical trajectory involving both the position and the flows.
The link with the studies of general time-additive observables 
via deformed Markov operators  \cite{derrida-lecture,sollich_review,lazarescu_companion,lazarescu_generic,jack_review,vivien_thesis,lecomte_chaotic,lecomte_thermo,lecomte_formalism,lecomte_glass,kristina1,kristina2,jack_ensemble,simon1,simon2,simon3,Gunter1,Gunter2,Gunter3,Gunter4,chetrite_canonical,chetrite_conditioned,chetrite_optimal,chetrite_HDR,touchette_circle,touchette_langevin,touchette_occ,touchette_occupation,derrida-conditioned,derrida-ring,bertin-conditioned,touchette-reflected,touchette-reflectedbis,c_lyapunov,previousquantum2.5doob,quantum2.5doob,quantum2.5dooblong,c_ruelle}
 can be then understood via the corresponding 'conditioned' process obtained from the generalization of Doob's h-transform.
 As in the field of stochastic resetting where the large deviations have been analyzed for excursions between resets
\cite{touchette2015,maj2019,c_reset}, it will be interesting to analyze also the large deviations for the empirical intervals between consecutive switching events of the run-and-tumble process.
 In the present paper, this will be called the Level 2.75 in order to stress that
it contains more information that the Level 2.5 that can be recovered via the contraction of the Level 2.75.

The paper is organized as follows.
In section \ref{sec_steady}, the one-dimensional run-and-tumble process with space-dependent velocities 
 and/or switching rates is described, and the condition for the existence of a localized non-equilibrium steady state
 is given.
The section \ref{sec_2.5} is devoted to the large deviations at Level 2.5 for the joint distribution of the empirical densities, of the empirical spatial currents and of the empirical switching flows.
This Level 2.5 is then contracted to obtain the Level 2 for the empirical densities alone in section \ref{sec_2},
and to characterize the large deviations of any time-additive observable in section \ref{sec_additive}.
In section \ref{sec_intervals}, the large deviations at Level 2.75 for the empirical intervals between consecutive switching events is analyzed.
Our conclusions are summarized in section \ref{sec_conclusion}.
The seven appendices contain various complementary computations and discussions with respect to the main text.


\section{ Non-uniform run-and-tumble processes with localized steady-states}

\label{sec_steady}

\subsection{ Run-and-tumble process with space-dependent velocities $v_{\pm}(x) >0$ and switching rates $ \gamma_{\pm}(x)\geq 0$ }

The particle is characterized by its position $x \in ]-\infty,+\infty[$ and by its internal state $\sigma=\pm$.
The dynamics can be summarized as follows :

(i) when at position $x$ and in the internal state $\sigma=+$, the particle moves with the positive velocity $v_+(x)>0$
unless the switching towards the other internal state $\sigma=-$ occurs with the switching rate $\gamma_{+}(x) \geq 0 $

(ii) when at position $x$ and in the internal state $\sigma=-$, the particle moves with the negative velocity $(-v_-(x))<0$
unless the switching towards the other internal state $\sigma=+$ occurs with the switching rate $\gamma_{-}(x) \geq 0 $

So the probabilities $P_{\sigma}( x,t)   $ to be at position $x$ and in the internal state $\sigma=\pm $ at time $t$
satisfy the dynamical system
\begin{eqnarray}
 \partial_t P_+( x,t)   &&   =  -    \partial_x      j_+(x,t)    -    Q_+(x,t) +  Q_-(x,t) 
\nonumber \\
 \partial_t P_-( x,t)  &&   =   -  \partial_x     j_-(x,t)   +   Q_+(x,t) -  Q_-(x,t) 
\label{run}
\end{eqnarray}
where the spatial currents $ j_{\pm}(x,t)  $ at position $x$ when the internal state is $\sigma=\pm 1$ at time $t$
involve the space-dependent velocities $v_{\pm}(x) >0$
\begin{eqnarray}
 j_+(x,t)  && \equiv  \ \  v_+( x ) P_+( x,t )  
\nonumber \\
   j_-(x,t)  && \equiv  -  v_-( x ) P_-( x,t )  
\label{spatialj}
\end{eqnarray}
while the switching flows $Q_{\pm}(x,t)  $ at position $x$ out of the internal state $\sigma=\pm 1$ at time $t$
(towards the opposite internal state $\sigma=\mp 1$ at the same position $x$) involve the space-dependent switching rates $ \gamma_{\pm}(x)\geq 0$
\begin{eqnarray}
 Q_+(x,t)  && \equiv   \gamma_+( x ) P_+( x,t )  
\nonumber \\
   Q_-(x,t)  && \equiv  \gamma_-( x ) P_-( x,t )  
\label{internalq}
\end{eqnarray}


\subsection{ Condition on the velocities $v_{\pm}(x) $ and on the switching rates $ \gamma_{\pm}(x)$ for a localized steady-state }

\label{subsec_steady}

Let us now consider the steady state of the dynamics of Eqs \ref{run}, 
with the two steady densities $P_{\pm}^*(x)$, 
 the two steady spatial currents
\begin{eqnarray}
 j^*_+(x)  && \equiv  \ \  v_+( x ) P_+^*( x)  
\nonumber \\
   j^*_-(x)  && \equiv  -  v_-( x ) P_-^*( x )  
\label{spatialjst}
\end{eqnarray}
and the two steady switching flows
\begin{eqnarray}
 Q^*_+(x)  && \equiv   \gamma_+( x ) P_+^*( x )  
\nonumber \\
   Q^*_-(x)  && \equiv  \gamma_-( x ) P_-^*( x )  
\label{internalqst}
\end{eqnarray}
that should satisfy
\begin{eqnarray}
0 && =  -    \frac{d  j^*_+(x)}{dx}    -  Q^*_+(x) + Q^*_-(x) 
\nonumber \\
0   && =   -    \frac{dj^*_-(x)}{dx}    + Q^*_+(x) - Q^*_-(x) 
\label{runstst}
\end{eqnarray}
The sum yields that the total steady-state spatial current 
\begin{eqnarray}
 j^*( x)     \equiv  j^*_+(x) +  j^*_-(x)  
\label{jxtot}
\end{eqnarray}
cannot depend on the position $x$,
and should thus vanish since no current is possible at $x \to \pm \infty$ 
\begin{eqnarray}
   0=j^*( x ) =   j^*_+(x) +  j^*_-(x)  
\label{continuityst}
\end{eqnarray}
In order to maintain a symmetric treatment of the two internal states $\sigma=\pm 1$,
it is more convenient to keep the spatial current $j^*_+(x)= -  j^*_-(x) $ as the fundamental observable,
from which the steady probabilities $P_{\pm}^*( x) $ can be obtained via Eqs \ref{spatialjst}
\begin{eqnarray}
P_+^*( x) && = \frac{ j^*_+(x) }{   v_+( x ) } 
\nonumber \\
P_-^*( x) && = \frac{ j^*_+(x) }{   v_-( x ) } 
\label{pfromjp}
\end{eqnarray}
and from which the steady switching flows can be obtained via Eqs \ref{internalqst}
\begin{eqnarray}
 Q^*_+(x)  && =   \gamma_+( x ) P_+^*( x )  =  \frac{ \gamma_+( x )}{   v_+( x ) }  j^*_+(x)
\nonumber \\
   Q^*_-(x)  && =  \gamma_-( x ) P_-^*( x )  = \frac{ \gamma_-( x )}{   v_-( x ) }  j^*_+(x)
\label{qfromjp}
\end{eqnarray}
Then Eq. \ref{runstst} yields the following closed equation for the current $  j^*_+(x)$
\begin{eqnarray}
 \frac{d  j^*_+(x)}{dx}  
  =  \left(  \frac{\gamma_-(x)}{v_-(x)} - \frac{\gamma_+(x)}{v_+(x)} \right)       j^*_+(x)
\label{eqjst}
\end{eqnarray}
The solution involving the integration constant $ j^*_+(0) $
\begin{eqnarray}
 j^*_+(x) =    j^*_+(0) \ e^{\displaystyle  \int_0^x dy \left[ \frac{ \gamma_-(y)    }{v_-(y) } - \frac{ \gamma_+(y) }{v_+( y)}  \right]}
\label{soljpst}
\end{eqnarray}
will be valid only if the corresponding steady state $ P^*_{\pm}( x )$ of Eq. \ref{pfromjp}
can be normalized
\begin{eqnarray}
1 && = \int_{-\infty}^{+\infty} dx \left[P^*_+( x ) +P^*_-( x) \right] 
= \int_{-\infty}^{+\infty} dx \left[  \frac{ 1 }{   v_+( x ) } + \frac{ 1 }{   v_-( x ) }  \right] j^*_+(x)
\nonumber \\
&&
= j^*_+(0) \int_{-\infty}^{+\infty} dx \left[  \frac{ 1 }{   v_+( x ) } + \frac{ 1 }{   v_-( x ) }  \right] 
 e^{\displaystyle  \int_0^x dy \left[ \frac{ \gamma_-(y)    }{v_-(y) } - \frac{ \gamma_+(y) }{v_+( y)}  \right]}
\label{stnorma}
\end{eqnarray}
 in order to determine the finite integration constant $ j^*_+(0) $ of Eq. \ref{soljpst}.

In conclusion, the two velocities $v_{\pm}(x) $ and the two switching rates $\gamma_{\pm}(x) $
 are able to produce a localized non-equilibrium steady-state only if the following integral is convergent
\begin{eqnarray}
\int_{-\infty}^{+\infty} dx \left[  \frac{ 1 }{   v_+( x ) } + \frac{ 1 }{   v_-( x ) }  \right] 
 e^{\displaystyle  \int_0^x dy \left[ \frac{ \gamma_-(y)    }{v_-(y) } - \frac{ \gamma_+(y) }{v_+( y)}  \right]}
 < +\infty
\label{stcv}
\end{eqnarray}

Appendix \ref{app_examples} contain
some simple examples of localized non-equilibrium steady states, where only the switching rates or only the velocities are space-dependent.


\section{ Large deviations at Level 2.5 for the empirical densities and flows }

\label{sec_2.5}

For a very long dynamical trajectory
$[x(0 \leq t \leq T); \sigma(0 \leq t \leq T)]$ of the run-and-tumble process of Eq. \ref{run} 
converging towards some localized non-equilibrium steady state (see the condition of Eq. \ref{stcv}),
the large deviations at Level 2.5 characterize the joint distribution
of the empirical time-averaged densities and empirical time-averaged flows as we now describe.

\subsection{ Empirical densities, empirical spatial currents and empirical switching flows with their constraints}

\label{subsec_empi}

The empirical densities $ \rho_{\sigma}( x)  $ measure the fraction of the time spent at position $x$ and in the internal state $\sigma=\pm$
\begin{eqnarray}
 \rho_{\sigma}( x) && \equiv \frac{1}{T} \int_0^T dt \  \delta_{\sigma(t),\sigma}  \ \delta (  x(t)-  x)  
\label{rhodiff}
\end{eqnarray}
with the normalization
\begin{eqnarray}
 \int_{-\infty}^{+\infty} dx \left[  \rho_+( x)  +  \rho_-( x) \right] =1
\label{rho1ptnormadiff}
\end{eqnarray}
The empirical spatial currents $ j_{\sigma}( x) $ at position $x$ while in the internal state $\sigma$
\begin{eqnarray} 
 j_{\sigma}( x) \equiv   \frac{1}{T} \int_0^T dt \  \frac{d  x(t)}{dt}   \delta(  x(t)-  x)  \ \delta_{\sigma(t),\sigma} \
\label{diffjlocaldef}
\end{eqnarray}
are completely determined by the empirical densities $\rho_{\sigma}( x) $ 
as a consequence of the deterministic motion at velocities $[\sigma v_{\sigma}(x) ]$ while in the internal state $\sigma$
\begin{eqnarray} 
 j_{+}( x) && =  v_+(x) \rho_{+}( x)  
 \nonumber \\
  j_{-}( x) && =  - v_-(x)  \rho_{-}( x)  
\label{diffjlocal}
\end{eqnarray}
The jumps between the two internal states $ \sigma=\pm$ at position $x$ are described by the two empirical switching flows
\begin{eqnarray}
Q_{+}( x)  \equiv  \frac{1}{T} \sum_{t  \in [0,T]: \sigma(t^+) \ne \sigma(t^-) } \delta_{\sigma(t^-),+} \  \delta(  x(t)-  x) 
\nonumber \\
Q_{-}( x)  \equiv  \frac{1}{T} \sum_{t  \in [0,T]: \sigma(t^+) \ne \sigma(t^-) } \delta_{\sigma(t^-),-} \  \delta(  x(t)-  x)
\label{jumpflows}
\end{eqnarray}
where the sum is over the finite number of times $t \in [0,T]$ where the trajectory $\sigma(t)$ is discontinuous 
$\sigma(t^+) \ne \sigma(t^-)$,
and jumps from the value $\sigma(t^-)=\sigma$ just before the jump at time $t^-$
towards the opposite value $\sigma(t^+)=-\sigma $ just after the jump at time $t^+$
(see section \ref{sec_intervals} for further details on the switching events
and for the more explicit forms of Eqs \ref{jumpflowsrecover} for the switching flows of Eqs \ref{jumpflows}).

The stationarity constraints read
\begin{eqnarray}
0 && = -   \frac{d j_+( x)  }{dx} - Q_{+}( x)  + Q_{-}( x)  
\nonumber \\
0 && = -   \frac{d j_-( x)  }{dx}+  Q_{+}( x)  - Q_{-}( x)  
\label{jumpstatio}
\end{eqnarray}
The sum yields that the total empirical spatial current 
\begin{eqnarray}
 j(x) && \equiv  j_{+}( x) +  j_{-}( x) = v_+(x) \rho_{+}( x)    - v_-(x)  \rho_{-}( x)  
\label{jxtotempi}
\end{eqnarray}
cannot depend on $x$, and should vanish as a consequence of the boundary conditions at $x \to \pm \infty$ with no flows
\begin{eqnarray}
0 =   j( x)  =    j_{+}( x) +  j_{-}( x)
\label{defjtot}
\end{eqnarray}
As for the steady state described in the previous section,
it will be more convenient to maintain a symmetric treatment of the two internal states $\sigma=\pm 1$,
and to keep the empirical spatial current $j_+(x)= -  j_-(x) $ as the fundamental observable,
from which the two empirical densities $\rho_{\pm}(x)$ can be computed 
\begin{eqnarray}
 \rho_+( x)   && = \frac{j_{+}( x)}{v_+(x) } 
 \nonumber \\
  \rho_-( x)   && = - \frac{j_{-}( x)}{v_-(x) }  = \frac{j_{+}( x)}{v_-(x) } 
\label{rhofromj}
\end{eqnarray}
while the remaining stationary constraint of Eq. \ref{jumpstatio} reads
\begin{eqnarray}
 \frac{d j_+( x)  }{dx} + Q_{+}( x)  - Q_{-}( x)  =0
\label{jumpstatiofinal}
\end{eqnarray}
The integral version of this stationary constraint reads
\begin{eqnarray}
  j_+( x) = \int_x^{+\infty} dy \left[Q_{+}( y)  - Q_{-}( y)  \right] = -  \int_{-\infty}^x dy \left[Q_{+}( y)  - Q_{-}( y)  \right] 
\label{jpintegral}
\end{eqnarray}
while the vanishing of the full integral
\begin{eqnarray}
 \int_{-\infty}^{+\infty} dy \left[Q_{+}( y)  - Q_{-}( y)  \right] = 0
\label{vanishing}
\end{eqnarray}
means that the total density of switching events out of the state $+$ and out of the state $-$ have to be equal.


\subsection{ Large deviations at Level 2.5 for the densities, the spatial currents and the switching flows  }

Taking into account the constitutive constraints of Eqs \ref{defjtot}
and
\ref{rhofromj}, one obtains
that the joint distribution of the empirical densities $\rho_{\pm}(x)$, of the empirical spatial currents $ j_{\pm}( x)  $
and the empirical switching flows $Q_{\pm}(x)$ can be factorized into 
\begin{eqnarray}
 P_T[ \rho_{\pm}(.), j_{\pm}(.) , Q_{\pm}(.) ]   \opsimeq_{T \to +\infty}   
\left[ \prod_{ x } \delta \left( \rho_+( x)   - \frac{j_{+}( x)}{v_+(x) }     \right) 
\delta \left( \rho_-( x)   - \frac{j_{+}( x)}{v_-(x) }     \right) \delta(j_-(x)+j_+(x) ) \right] 
P_T[  j_+(.) , Q_{\pm}(.) ]  
\label{ld2.5rhoqfull}
\end{eqnarray}
where the joint distribution $P_T[  j_+(.) , Q_{\pm}(.) ]  $ of the three remaining variables satisfy the large deviation form
\begin{eqnarray}
 P_T[  j_+(.) , Q_{\pm}(.) ]   \opsimeq_{T \to +\infty} 
&&  \delta \left( \int_{-\infty}^{+\infty} dx \left[  \frac{1}{v_+(x) }   +  \frac{1}{v_-(x) } \right] j_+(x)   - 1  \right)
 \left[ \prod_{ x } \delta \left( Q_{+}( x)  - Q_{-}( x)  
+ \frac{d j_+(x) }{dx}    \right)  \right] 
   \nonumber \\
&& e^{- \displaystyle T  I_{2.5}[j_+(.),  Q_{\pm}(.) ]   }
\label{ld2.5rhoq}
\end{eqnarray}
The constraints of the first line correspond to the normalization of Eq. \ref{rho1ptnormadiff}
and to the stationarity constraint of Eq. \ref{jumpstatiofinal}, while the rate function at Level 2.5
follows the standard form for Markov jump processes
\cite{fortelle_thesis,fortelle_jump,maes_canonical,maes_onandbeyond,wynants_thesis,chetrite_formal,BFG1,BFG2,chetrite_HDR,c_ring,c_interactions,c_open,c_detailed,barato_periodic,chetrite_periodic,c_reset,c_inference}
\begin{eqnarray}
 I_{2.5}[ j_+,  Q_{\pm}(.) ]  && = 
  \int_{-\infty}^{+\infty}d  x
\left[  Q_{+}( x)    \ln \left( \frac{  Q_{+}( x)   }{     \frac{\gamma_+(x)}{v_+(x) }  j_+(x) }  \right) 
 -   Q_{+}( x)   +    \frac{\gamma_+(x)}{v_+(x) }  j_+(x)    \right]  
\nonumber \\
&& +
  \int_{-\infty}^{+\infty}d  x
\left[  Q_{-}( x)   \ln \left( \frac{  Q_{-}( x)   }{    \frac{\gamma_-(x)}{v_-(x) }  j_+(x) }  \right) 
 -   Q_{-}( x)   +   \frac{\gamma_-(x)}{v_-(x) }  j_+(x)  \right]  
\label{rate2.5rhoq}
\end{eqnarray}

Appendix \ref{app_alternative} contains the inference interpretation of the Level 2.5,
while the two next sections describe how the Level 2.5 can be contracted to obtain 
other large deviations properties.


\section{ From the Level 2.5 towards the Level 2 for the empirical densities alone }

\label{sec_2}

As recalled in the Introduction, the Level 2 for the empirical densities alone is not closed for 
non-equilibrium steady states involving steady currents, so that it can only be obtained 
via the optimization of the explicit Level 2.5 (described in the previous section)
over the empirical flows. 
It is convenient to make this contraction in two steps via the introduction of the intermediate Level 2.25.

\subsection{ Level 2.25 for the distribution of the spatial current $j_+(x)$ and of the switching current $J( x)
$ }

As explained in Appendix \ref{app_contractionfrom2.5to2.25},
the explicit contraction over the switching activity $A(x) \equiv Q_{+}( x)  + Q_{-}( x)  $
yields that
the joint distribution of the spatial current $j_+(x)$ and of the switching current $J(x) \equiv Q_{+}( x)  - Q_{-}( x)  $
follows the large deviation form
\begin{eqnarray}
 P_T[  j_+(.), J(.)]   \opsimeq_{T \to +\infty}  
  \delta \left( \int_{-\infty}^{+\infty}  dx \left[  \frac{1}{v_+(x) }   +  \frac{1}{v_-(x) } \right] j_+(x)   - 1  \right)
 \left[ \prod_{ x } \delta \left( J( x)  + \frac{d j_+(x) }{dx}    \right)  \right]  
  e^{ \displaystyle -   T I_{2.25} [ j_+(.),  J(.)]   }
\label{ld2.5diffwithoutA}
\end{eqnarray}
where the rate function $ I_{2.25} [ j_+(.),  J(.)] $ at Level 2.25 reads 
\begin{eqnarray}
 I_{2.25} [ j_+(.),  J(.)] 
&& =
  \int_{-\infty}^{+\infty}d  x
  J( x)   \ln \left( \frac{\sqrt{   J^2( x)  + 4 \frac{\gamma_+(x)\gamma_-(x)}{v_+(x)v_-(x)}   j_+^2( x)  }   +J(x)  }
  {  2  \frac{\gamma_+(x)}{v_+(x) }  j_+(x)    }  \right)
 \nonumber \\ &&
 +  \int_{-\infty}^{+\infty}d  x
  \left[\left( \frac{\gamma_+(x)}{v_+(x) }    +   \frac{\gamma_-(x)}{v_-(x) } \right) j_+(x) 
  -   \sqrt{   J^2( x)  + 4  \frac{\gamma_+(x)\gamma_-(x)}{v_+(x)v_-(x)}   j_+^2( x)  }   
  \right]
\label{rate2.25}
\end{eqnarray}


\subsection{ Large deviations at Level 2 for the empirical densities $ \rho_{\pm}( x)$ alone }

The distribution $P_T[  j_+(.)  ]  $ of the spatial current $j_+(x)$ alone 
can be obtained from Eq. \ref{ld2.5diffwithoutA}
via the elimination of the switching current $J(x)$ in terms of the spatial current $j_+(x)$
\begin{eqnarray}
J(x) = -   j_+'(x)  
\label{jconstraintrho}
\end{eqnarray}
and reads
\begin{eqnarray}
 P_T[ j_+(.)]   \opsimeq_{T \to +\infty}  
   \delta \left( \int_{-\infty}^{+\infty} dx \left[  \frac{1}{v_+(x) }   +  \frac{1}{v_-(x) } \right] j_+(x)   - 1  \right) 
   e^{ \displaystyle -   T I_2 [ j_+(.)]   }
\label{ld2rhoalone}
\end{eqnarray}
with the rate function
\begin{eqnarray}
 I_2 [ j_+(.)] &&  =
-   \int_{-\infty}^{+\infty}d  x
    j_+'(x)     \ln \left( \frac{\sqrt{  [j_+'(x)]^2  + 4 \frac{\gamma_+(x)\gamma_-(x)}{v_+(x)v_-(x)}   j_+^2( x)  }   -   j_+'(x)    }
  {  2  \frac{\gamma_+(x)}{v_+(x) }  j_+(x)    }  \right)
 \nonumber \\ &&
 +  \int_{-\infty}^{+\infty}d  x
  \left(
    \left[  \frac{\gamma_+(x)}{v_+(x) }   +  \frac{\gamma_-(x)}{v_-(x) } \right] j_+(x)
  -   \sqrt{   [j_+'(x)]^2    + 4    \frac{\gamma_+(x)\gamma_-(x)}{v_+(x)v_-(x)}   j_+^2( x) } 
  \right)
  \label{i2rhoalone}
\end{eqnarray}
Using Eq. \ref{ld2.5rhoqfull}, one obtains that the Level 2
for the joint distribution of the two densities $ \rho_{+}( x)=\frac{j_{+}( x)}{v_+(x) }$  and  
$ \rho_{-}( x)=\frac{j_{+}( x)}{v_-(x) }$
reads
\begin{eqnarray}
 P_T\left[ \rho_{+}(.)= \frac{j_{+}( .)}{v_+(.)} ; \rho_{-}( .)=\frac{j_{+}( .)}{v_-(.) } \right]   \opsimeq_{T \to +\infty}  
   \delta \left( \int_{-\infty}^{+\infty} dx \left[  \frac{1}{v_+(x) }   +  \frac{1}{v_-(x) } \right] j_+(x)   - 1  \right) 
   e^{ \displaystyle -   T I_2 [ j_+(.)]   }
\label{Level2}
\end{eqnarray}

This Level 2 could be further contracted to obtain the Level 1 concerning 
time-addtitive observables of the position only. However, it is more interesting to consider
the case of more general time-additive observables that involve also the switching flows as
explained in the next section.


\section{ Large deviations of time-additive observables via contraction of Level 2.5 }

\label{sec_additive}

In this section, the goal is to analyze the large deviations properties of
 the most general time-additive observable of the run-and-tumble trajectory $[x(0 \leq t \leq T); \sigma(0 \leq t \leq T)]$
 that can a priori be parametrized by the six functions $(\alpha_{\pm}(x);\beta_{\pm}(x);\omega_{\pm}(x))$
\begin{eqnarray}
 A_T   \equiv   \frac{1}{T}\int_0^T dt  \left[ \alpha_{\sigma(t)}(  x(t)) + \dot x(t) \omega_{\sigma(t)}(  x(t))\right]
+  \frac{1}{T} \sum_{t  \in [0,T] : \sigma(t^+) \ne \sigma(t^-) } \beta_{\sigma(t^-)}(x(t) )
\label{additiveAdiff}
\end{eqnarray}


\subsection { Time-additive observable $A_T$ in terms of the empirical densities and of the empirical flows }

The additive observable of Eq. \ref{additiveAdiff}
can be rewritten in terms of the empirical densities $\rho_{\pm}(x)$ of Eq. \ref{rhodiff},
in terms of the empirical spatial currents $j_{\pm}(x) $ of Eq. \ref{diffjlocaldef}
and in terms of the empirical switching flows $ Q_{\pm}( x)$ of Eq. \ref{jumpflows} as
\begin{eqnarray}
 A_T  = \int_{-\infty}^{+\infty}d  x \sum_{\sigma=\pm 1}
 \left[ \alpha_{\sigma}(  x)  \rho_{\sigma}(  x ) + \omega_{\sigma}(  x)  j_{\sigma}(  x ) 
 + \beta_{\sigma}(x) Q_{\sigma}(x)  \right]
\label{additiveAempi}
\end{eqnarray}
As discussed in detail in the subsection \ref{subsec_empi},
these empirical observables are not all independent, so that the parametrization with six functions is 
actually very redundant. Rewriting $j_-(x)$ and $\rho_{\pm}(x)$ in terms of $j_+(x)$ via Eqs \ref{defjtot} and \ref{rhofromj},
one obtains that the additive observable of Eq. \ref{additiveAempi} can be parametrized in terms of three functions 
$(\omega(x),\beta_{\pm}(x))$
\begin{eqnarray}
 A_T  = \int_{-\infty}^{+\infty}d  x 
 \left[  \omega(x)    j_+(x)
 + \beta_{+}(x) Q_{+}(x) + \beta_{-}(x) Q_{-}(x) \right]
\label{additiveAempijp}
\end{eqnarray}
where the new function $\omega(x)$ takes into account the four previous functions $(\omega_{\pm}(x),\alpha_{\pm}(x))$
\begin{eqnarray}
 \omega(x)   \equiv  \omega_+(x) -  \omega_-(x) + \frac{\alpha_{+}( x)}{v_+(x) } + \frac{\alpha_{-}( x)}{v_-(x) } 
\label{omega}
\end{eqnarray}


\subsection { Analysis of the generation function of $A_T$ via the contraction of the Level 2.5 }

The rewriting of Eq. \ref{additiveAempijp} yields that the generating function of the additive observable $A_T$
 \begin{eqnarray}
 Z_T(k) \equiv  &&   < e^{ \displaystyle T k A_T } >  
 \nonumber \\&& 
=
\int {\cal D} j_+(.) \int {\cal D} Q_{\pm}(.)  P_T[  j_+(.) , Q_{\pm}(.) ] 
e^{\displaystyle T k \int_{-\infty}^{+\infty}d  x 
 \left[  \omega(x)    j_+(x) + \beta_{+}(x) Q_{+}(x) + \beta_{-}(x) Q_{-}(x) \right]}
\label{genek}
\end{eqnarray}
can be evaluated from the joint distribution $ P_T[  j_+(.) , Q_{\pm}(.) ]  $ of Eq. \ref{ld2.5rhoq}
and Eq. \ref{rate2.5rhoq}.
As a consequence, the asymptotic behavior of the generating function of Eq. \ref{genek} for large $T$
 \begin{eqnarray}
 Z_T(k)   \opsimeq_{T \to +\infty} 
&& \int {\cal D} j_+(.) \int {\cal D} Q_{\pm}(.)  
 \delta \left( \int_{-\infty}^{+\infty} dx \left[ \frac{1}{v_+(x) } + \frac{1}{v_-(x) }   \right] j_+(x)   - 1  \right) 
\left[ \prod_{ x } \delta \left( Q_{+}( x)  - Q_{-}( x)  + \frac{d j_+(x) }{dx}    \right)  \right] 
   \nonumber \\ && 
 e^{ \displaystyle T \left( - I_{2.5}[j_+(.),  Q_{\pm}(.) ]   + k \int_{-\infty}^{+\infty}d  x 
 \left[  \omega(x)    j_+(x) + \beta_{+}(x) Q_{+}(x) + \beta_{-}(x) Q_{-}(x) \right] \right)}
\label{genekcol}
\end{eqnarray}
can be evaluated via the saddle-point method : one needs to optimize 
the functional appearing in the exponential of the second line 
 in the presence of the constraints of the first line.


\subsection { Optimization of the appropriate Lagrangian $ {\cal L}_k  [j_+(.),  Q_{\pm}(.) ]  $ }

 In order to solve the optimization problem of Eq. \ref{genekcol}, 
 let us introduce the following Lagrangian with
the Lagrange multipliers $(\mu(k) ,\lambda_k(.))$ to take into account the constraints
 \begin{eqnarray}
&& {\cal L}_k  [j_+(.),  Q_{\pm}(.) ] 
 \equiv  - I_{2.5}[j_+(.),  Q_{\pm}(.) ]   + k \int_{-\infty}^{+\infty}d  x 
 \left[  \omega(x)    j_+(x) + \beta_{+}(x) Q_{+}(x) + \beta_{-}(x) Q_{-}(x) \right]
\nonumber \\
&& - \mu(k) \left( \int_{-\infty}^{+\infty} dx \left[ \frac{1}{v_+(x) } + \frac{1}{v_-(x) }   \right] j_+(x)   - 1  \right) 
-  \int_{-\infty}^{+\infty}d  x \lambda_k(x)  \left( Q_{+}( x)  - Q_{-}( x)  + \frac{d j_+(x) }{dx}    \right) 
\label{lagrange}
\end{eqnarray}
Using the explicit form of the rate function of Eq. \ref{rate2.5rhoq} and performing an integration by part 
of the last term involving the derivative $\frac{d j_+(x) }{dx} $, this Lagrangian can be rewritten 
more compactly as
 \begin{eqnarray}
{\cal L}_k  [j_+(.),  Q_{\pm}(.) ] 
 \equiv  \mu(k)
&& +  \int_{-\infty}^{+\infty}d  x
\left[ - Q_{+}( x)    \ln \left( \frac{  Q_{+}( x)   }{     \frac{ \gamma_{+}( x)}{v_+(x) } e^{k \beta_+(x)- \lambda_k(x)} j_{+}( x)}  \right) 
 +  Q_{+}( x)     \right]  
\nonumber \\
&& +
  \int_{-\infty}^{+\infty}d  x
\left[ - Q_{-}( x)   \ln \left( \frac{  Q_{-}( x)   }{     \frac{ \gamma_{-}( x)}{v_-(x) } e^{k \beta_-(x)+ \lambda_k(x) }j_{+}( x) } \right) 
 +   Q_{-}( x)   \right]  
\nonumber \\  
&& +  \int_{-\infty}^{+\infty}d  x 
 \left[ k  \omega(x) -    \frac{ \gamma_{+}( x)+ \mu(k)}{v_+(x) }  -   \frac{  \gamma_{-}( x)+ \mu(k)}{v_-(x) }  
 +  \lambda_k'(x) \right]  j_+(x) 
\label{lagrange2}
\end{eqnarray}
The optimization with respect to $Q_{+}(x)$, $Q_-(x)$ and $j_+(x)$ read
 \begin{eqnarray}
0 && = \frac{ \partial {\cal L}_k  [j_+(.),  Q_{\pm}(.) ] }{\partial Q_+(x) }
 =   -     \ln \left( \frac{  Q_{+}( x)   }{     \frac{ \gamma_{+}( x)}{v_+(x) } e^{k \beta_+(x)- \lambda_k(x)  } j_{+}( x)}  \right)  
 \nonumber \\
 0 && = \frac{ \partial {\cal L}_k  [j_+(.),  Q_{\pm}(.) ] }{\partial Q_-(x) }
= -    \ln \left( \frac{  Q_{-}( x)   }{     \frac{ \gamma_{-}( x)}{v_-(x) } e^{k \beta_-(x)+   \lambda_k(x) }j_{+}( x) } \right)   
 \nonumber \\
 0 && = \frac{ \partial {\cal L}_k  [j_+(.),  Q_{\pm}(.) ] }{\partial j_+(x) }
 = \frac{Q_+(x)+Q_-(x)}{j_+(x)} + k  \omega(x) -    \frac{ \gamma_{+}( x)+ \mu(k)}{v_+(x) }  -   \frac{  \gamma_{-}( x)+ \mu(k)}{v_-(x) }  
 +  \lambda_k'(x)
\label{lagrangederi}
\end{eqnarray}
Plugging the optimal solutions of the two first equations 
 \begin{eqnarray}
   Q^{+opt}_{k}( x) &&  =    \frac{ \gamma_{+}( x)}{v_+(x) } e^{k \beta_+(x) - \lambda_k(x) } j^{+opt}_{k}( x) 
 \nonumber \\
  Q^{-opt}_{k}( x) &&  =    \frac{ \gamma_{-}( x)}{v_-(x) } e^{k \beta_-(x)+\lambda_k(x)}j^{+opt}_{k}( x) 
\label{qopt}
\end{eqnarray}
into the third equation yields the first-order non-linear 
differential equation for the Lagrange multiplier $ \lambda_k(x)$
 \begin{eqnarray}
 0 
 = \frac{ \gamma_{+}( x) e^{k \beta_+(x) - \lambda_k(x) } - \gamma_{+}( x)- \mu(k)}{v_+(x) } 
 + \frac{ \gamma_{-}( x) e^{k \beta_-(x)+\lambda_k(x)} - \gamma_{-}( x)- \mu(k)}{v_-(x) } 
  + k  \omega(x)     +  \lambda_k'(x)
\label{lagrangederi3}
\end{eqnarray}
In addition, the optimal solutions of Eq. \ref{qopt} should satisfy the constraints of the last line of Eq. \ref{lagrange}
 \begin{eqnarray}
1 && = \int_{-\infty}^{+\infty} dx \left[ \frac{1}{v_+(x) } + \frac{1}{v_-(x) }   \right] j^{+opt}_{k}(x)
\nonumber \\
\frac{d j^{+opt}_{k}(x) }{dx}&& = - Q^{+opt}_{k}( x)  + Q^{-opt}_{k}( x)  
= \left(  - \frac{ \gamma_{+}( x)}{v_+(x) } e^{k \beta_+(x) - \lambda_k(x) } 
 +    \frac{ \gamma_{-}( x)}{v_-(x) } e^{k \beta_-(x)+\lambda_k(x)} \right) j^{+opt}_{k}( x) 
\label{lagrangec}
\end{eqnarray}
Finally, one needs to evaluate the optimal value of the Largrangian of Eq. \ref{lagrange2} 
and one obtains that it reduces to the Lagrange multiplier $\mu(k)$
 \begin{eqnarray}
{\cal L}_k^{opt} \equiv  {\cal L}_k  [j^{+opt}_{k}(.),  Q^{\pm opt}_{k}(.) ] = \mu(k)
\label{lagrangeopt}
\end{eqnarray}
So the asymptotic behavior of the generating function of Eq. \ref{genekcol}
 \begin{eqnarray}
 Z_T(k)   \opsimeq_{T \to +\infty} 
e^{ \displaystyle T {\cal L}_k^{opt} } = e^{ \displaystyle T \mu(k)}
\label{genekcolmu}
\end{eqnarray}
yields that the Lagrange multiplier $\mu(k)$ associated to the normalization constraint
is the scaled cumulant generating function of the time-additive variable $A_T$ as already found in other problems \cite{c_reset,c_lyapunov}.


\subsection { Summary of the optimization procedure in three steps }

\label{subsec_optimization3steps}

The optimization procedure described above can be thus decomposed in three steps :

(1) the scaled cumulant generating function $\mu(k)$
and the Lagrange multiplier $\lambda_k(x)$ have to be found together from 
 the differential equation for the Lagrange multiplier $ \lambda_k(x)$
 of Eq. \ref{lagrangederi3} that can be rewritten as
 \begin{eqnarray}
   \lambda_k'(x) 
   + \left(\frac{ \gamma_{+}( x) e^{k \beta_+(x)  } }{v_+(x) } \right) e^{- \lambda_k(x) } 
 + \left( \frac{ \gamma_{-}( x) e^{k \beta_-(x)} }{v_-(x) } \right) e^{ \lambda_k(x) } 
  = 
   \frac{  \gamma_{+}( x)+ \mu(k)}{v_+(x) } 
 + \frac{  \gamma_{-}( x)+ \mu(k)}{v_-(x) } 
  -  k  \omega(x)      
\label{lagrangederi3expli}
\end{eqnarray}
 If one is only interested into the scaled cumulant generating function $\mu(k)$, one can actually stop here,
 while if one is also interested into the saddle-point solution, one goes on with the two other steps.

(2)  the optimal current $j^{+opt}_{k}( x)$ determined by Eqs \ref{lagrangec}
is then given by the analog of Eqs \ref{soljpst} and \ref{stnorma}
 \begin{eqnarray}
 j^{+opt}_{k}(x) =  
\frac{ e^{ \displaystyle  \int_0^x dy \left(\frac{ \gamma_{-}( y)}{v_-(y) } e^{k \beta_-(y)+\lambda_k(y)} 
  - \frac{ \gamma_{+}( y)}{v_+(y) } e^{k \beta_+(y) - \lambda_k(y) }  \right) } }
  {  \int_{-\infty}^{+\infty} dz \left[ \frac{1}{v_+(z) } + \frac{1}{v_-(z) }   \right] 
   e^{ \displaystyle  \int_0^z dy \left(\frac{ \gamma_{-}( y)}{v_-(y) } e^{k \beta_-(y)+\lambda_k(y)} 
  - \frac{ \gamma_{+}( y)}{v_+(y) } e^{k \beta_+(y) - \lambda_k(y) }  \right) }}
\label{joptsolu}
\end{eqnarray}
that involves the Lagrange multiplier $ \lambda_k(x)$ found in step (1)

(3) the optimal flows $Q^{\pm opt}_{k}( x)  $ are then given by Eqs \ref{qopt}
using the Lagrange multiplier $ \lambda_k(x)$ found in step (1) and the optimal current $j^{+opt}_{k}( x) $ found in step (2)
 \begin{eqnarray}
   Q^{+opt}_{k}( x) &&  =    \frac{ \gamma_{+}( x)}{v_+(x) } e^{k \beta_+(x) - \lambda_k(x) } j^{+opt}_{k}( x) 
 \nonumber \\
  Q^{-opt}_{k}( x) &&  =    \frac{ \gamma_{-}( x)}{v_-(x) } e^{k \beta_-(x)+\lambda_k(x)}j^{+opt}_{k}( x) 
\label{qoptstep3}
\end{eqnarray}

Appendix \ref{app_perturbativeoptimization} describes the perturbative solution in the parameter $k$
of Eq. \ref{lagrangederi3expli} in order to obtain the two first cumulants of any time-additive observable $A_T$.
The equivalence 
of this optimization procedure with the standard deformed generator method to analyze the large deviations of 
additive observables is explained in Appendix \ref{app_doob}.


\section{ Large deviations at Level 2.75 for the intervals between switching events }

\label{sec_intervals}

Up to now, we have described the Level 2.5 in section \ref{sec_2.5} and its contractions in sections 
\ref{sec_2} and \ref{sec_additive}. In the present section, we consider instead the higher Level 2.75
concerning the empirical intervals between consecutive switching events.

\subsection{ Decomposition of a long trajectory into its intervals between consecutive switching events}

For a very long trajectory of initial internal state $\sigma(t=0)=-$, let us introduce
 the times $t_i$ with $i=1,..,2N-1$ where the internal state $\sigma(t)$ changes.
It is convenient to add $t_0=0$ for the initial time and $t_{2N}=T$ for the end of the trajectory.
During the odd intervals $t_{2i}  < t < t_{2i+1}$, the internal state is negative $\sigma(t)=-$ 
and the spatial trajectory $x(t)$ is ballistic with the negative velocity $(-v_-(x)<0)$.
During the even intervals $t_{2i-1}  < t < t_{2i}$, the internal state is positive $\sigma(t)=+$
and the spatial trajectory $x(t)$ is ballistic with the positive velocity $(v_+(x)>0)$.
Since the velocities $v_{\pm}(x)$ and the switching rates $\gamma_{\pm}(x)$ are space-dependent, 
it is convenient to characterize each interval by 
its left extremal position $x_L$ and by its right extremal position $x_R$.

For an odd interval starting at position $x_R$,
the probability to end at the position $x_L$ reads
\begin{eqnarray}
W_-(x_L \vert x_R) 
= \frac{\gamma_-(x_L) }{v_-(x_L)} \ e^{\displaystyle - \int_{x_L}^{x_R} dx \frac{\gamma_-( x )}{v_-(x)}  }
= \frac {\partial } {\partial x_L}    \ \ e^{\displaystyle - \int_{x_L}^{x_R} dx \frac{\gamma_-( x )}{v_-(x)}  }
\label{emoins}
\end{eqnarray}
with the normalization
\begin{eqnarray}
\int_{-\infty}^{x_R} dx_L W_-(x_L \vert x_R) =1
\label{emoinsn}
\end{eqnarray}
The time duration associated to an odd interval of extremal positions $(x_L,x_R)$ involves the velocity $v_-(x)$
on this interval
\begin{eqnarray}
\tau_{-}(x_L,x_R) = \int_{x_L}^{x_R} \frac{dx}{v_{-}(x)} 
\label{durationsinterm}
\end{eqnarray}

Similarly, for an even interval starting at position $x_L$,
the probability to end at the position $x_R$ reads
\begin{eqnarray}
W_+(x_R \vert x_L) 
= \frac{\gamma_+(x_R) }{v_+(x_R)} \ e^{\displaystyle - \int_{x_L}^{x_R} dx \frac{\gamma_+( x )}{v_+(x)}  }
=  - \frac {\partial } {\partial x_R}    \ \ e^{\displaystyle - \int_{x_L}^{x_R} dx \frac{\gamma_+( x )}{v_+(x)}  }
\label{eplus}
\end{eqnarray}
with the normalization
\begin{eqnarray}
\int_{x_L}^{+\infty} dx_R W_+(x_R \vert x_L) =1
\label{eplusn}
\end{eqnarray}
The time duration associated to an even interval of extremal positions $(x_L,x_R)$ involves the velocity $v_+(x)$
on this interval
\begin{eqnarray}
\tau_{+}(x_L,x_R) = \int_{x_L}^{x_R} \frac{dx}{v_{+}(x)} 
\label{durationsinterp}
\end{eqnarray}

So if one considers only the spatial positions $x(t_j)$ at the times $t_j$ where the internal state $\sigma(t)$ changes,
the probabilities $P_{t_{2i}}(x(t_{2i})=x_R)$ and $ P_{t_{2i+1}}(x(t_{2i+1})=x_L)$
follow the alternate Markov chain
\begin{eqnarray}
P_{t_{2i}}(x_R) && = \int_{-\infty}^{x_R} dx_L W_+(x_R \vert x_L) P_{t_{2i-1}}(x_L) 
\delta \left( t_{2i} - t_{2i-1} - \tau_{+}(x_L,x_R)\right)
\nonumber \\
P_{t_{2i+1}}(x_L) && = \int_{x_L}^{+\infty} dx_R W_-(x_L \vert x_R) P_{t_{2i}}(x_R) 
\delta \left( t_{2i+1} - t_{2i} - \tau_{-}(x_L,x_R)\right)
\label{alternatemarkov}
\end{eqnarray}


\subsection{ Empirical densities of intervals between switching events with their constraints }

The empirical densities of intervals between consecutive switching events 
\begin{eqnarray}
q_{+} (x_R,x_L) && \equiv \frac{1}{T} \sum_{i=1}^N  \delta( x_R-  x(t_{2i}) ) \delta( x_L- x(t_{2i-1}))
\nonumber \\
q_{-} (x_L,x_R) && \equiv \frac{1}{T} \sum_{i=0}^{N-1}  \delta( x_L-  x(t_{2i+1}) )  \delta(x_R- x(t_{2i}))
\label{Q2}
\end{eqnarray}
are defined for $x_L < x_R$
and contain the information of the empirical switching flows $Q_{\pm}(x)$ of Eq. \ref{jumpflows} that can be rewritten as
\begin{eqnarray}
Q_{+}( x_R)  && = \frac{1}{T} \sum_{i=1}^N  \delta( x_R-  x(t_{2i}) )
=   \int_{-\infty}^{x_R} dx_L q_{+} (x_R,x_L)
=  \int_{-\infty}^{x_R} dx_L q_{-} (x_L,x_R)
\nonumber \\
Q_{-}( x_L) && = \frac{1}{T} \sum_{i=1}^N     \delta( x_L- x(t_{2i-1}))
=  \int_{x_L}^{+\infty} dx_R q_{+} (x_R,x_L)
=  \int_{x_L}^{+\infty} dx_R q_{-} (x_L,x_R)
\label{jumpflowsrecover}
\end{eqnarray}
Their common normalization corresponds to the density $n=\frac{N}{T} $ of positive or negative intervals (Eq. \ref{vanishing})
\begin{eqnarray}
n=\frac{N}{T} && =\int_{-\infty}^{+\infty} dx_R \ Q_+ (x_R)  = \int_{-\infty}^{+\infty} dx_L \ Q_-(x_L) 
\nonumber \\
&& =  \int_{-\infty}^{+\infty} dx_L \int_{x_L}^{+\infty} dx_R q_{+} (x_R,x_L)
= \int_{-\infty}^{+\infty} dx_L \int_{x_L}^{+\infty} dx_R q_{-} (x_R,x_L)
\label{n1empit}
\end{eqnarray}
The empirical densities of intervals $q_{\pm}(.,.) $ of Eq. \ref{Q2} also contain the information on the empirical densities of Eq. \ref{rhodiff}
\begin{eqnarray}
\rho_{+}(x)   
&& =   \frac{ 1 }{v_+(x)}  \int_{-\infty}^{x} dx_L \int_{x}^{+\infty} dx_R  q_{+} (x_R,x_L) 
\nonumber \\
\rho_{-}(x) 
 && =  \frac{ 1 }{v_-(x)}  \int_{-\infty}^{x} dx_L \int_{x}^{+\infty} dx_R   q_{-} (x_L,x_R) 
\label{recoverrho}
\end{eqnarray}
or equivalently on the empirical currents of Eq. \ref{diffjlocal}
\begin{eqnarray} 
 j_{+}( x) && =  v_+(x) \rho_{+}( x)  =\int_{-\infty}^{x} dx_L \int_{x}^{+\infty} dx_R  q_{+} (x_R,x_L) 
 \nonumber \\
  j_{-}( x) && =  - v_-(x)  \rho_{-}( x)  = - \int_{-\infty}^{x} dx_L \int_{x}^{+\infty} dx_R   q_{-} (x_L,x_R) 
\label{diffjlocalinter}
\end{eqnarray}
whose derivatives 
can be rewritten in terms of $Q_{\pm}$ using Eqs \ref{jumpflowsrecover}
\begin{eqnarray}
 j_{+}'( x)
&& =  \int_{x}^{+\infty} dx_R    q_{+} (x_R,x) 
-  \int_{-\infty}^{x} dx_L     q_{+} (x,x_L) 
= Q_-(x)-Q_+(x)
\nonumber \\
 - j_{-}'( x)
 && =   \int_{x}^{+\infty} dx_R    q_{-} (x,x_R) 
 -  \int_{-\infty}^{x} dx_L     q_{-} (x_L,x) 
 =  Q_-(x)-Q_+(x)
\label{recoverrhoderi}
\end{eqnarray}
so that one recovers the stationarity constraint of Eq. \ref{jumpstatiofinal}.

The normalization of Eq. \ref{rho1ptnormadiff} for the total empirical density
 using Eq. \ref{recoverrho}
\begin{eqnarray}
1 && = \int_{-\infty}^{+\infty} dx \left[  \rho_+( x)  +  \rho_-( x) \right] 
= \int_{-\infty}^{+\infty} dx \left[  \frac{1}{v_+(x) }   +  \frac{1}{v_-(x) } \right] j_+(x)
\nonumber \\
&& =
\int_{-\infty}^{+\infty} dx_L \int_{x_L}^{+\infty} dx_R  
\left[ q_{+} (x_R,x_L) \int_{x_L}^{x_R} \frac{dx}{v_+(x)} 
+  q_{-} (x_R,x_L)\int_{x_L}^{x_R} \frac{dx}{v_-(x)} \right]
\nonumber \\
&& =
\int_{-\infty}^{+\infty} dx_L \int_{x_L}^{+\infty} dx_R  
\left[ q_{+} (x_R,x_L) \tau_{+}(x_L,x_R) 
+  q_{-} (x_R,x_L) \tau_{-}(x_L,x_R) \right]
\label{normarhointer}
\end{eqnarray}
 involves the time durations $\tau_{\pm}(x_L,x_R) $ of Eqs \ref{durationsinterm} and \ref{durationsinterp}
 of the two types of intervals.
 So Eq. \ref{normarhointer}
  corresponds to the normalization of the total time $T$ of the trajectory
  when decomposed into the durations of all 
  empirical intervals.


\subsection{ Large deviations at Level 2.75 for the empirical densities of intervals between switching events }

The joint distribution of the empirical current $j_+(.)$, of the empirical switching flows $Q_{\pm}(.) ] $
and of the empirical densities of intervals $q_{\pm} (.,.)$ 
follows the large deviation form
\begin{eqnarray}
&& P_T[   j_+(.), Q_{\pm}(.), q_{\pm} (.,.)   ] \opsimeq_{T \to +\infty}  C_{2.75}[   j_+(.),  Q_{\pm}(.), q_{\pm} (.,.)   ] 
 e^{\displaystyle - T  I_{2.75} [ j_+(.),  Q_{\pm}(.), q_{\pm} (.,.)  ]    }
\label{ld2.75}
\end{eqnarray}
The constraints
\begin{eqnarray}
 && C_{2.75}[   j_+(.),  Q_{\pm}(.), q_{\pm} (.,.)   ]  =
  \delta \left( \int_{-\infty}^{+\infty} dx \left[  \frac{1}{v_+(x) }   +  \frac{1}{v_-(x) } \right] j_+(x)   - 1  \right)
   \left[ \prod_{ x } \delta \left( Q_{+}( x)  - Q_{-}( x)  
+ \frac{d j_+(x) }{dx}    \right)  \right]
\nonumber \\
&&\left[ \prod_{ x_R } \delta \left(  \int_{-\infty}^{x_R} dx_L q_{+} (x_R,x_L)  - Q_{+}( x_R)  \right)  \right]
\left[ \prod_{ x_R } \delta \left( \int_{-\infty}^{x_R} dx_L q_{-} (x_L,x_R) - Q_{+}( x_R)   \right)  \right]
\nonumber \\
&&\left[ \prod_{ x_L } \delta \left(   \int_{x_L}^{+\infty} dx_R q_{+} (x_R,x_L) - Q_{-}( x_L)   \right)  \right]
\left[ \prod_{ x_L } \delta \left(   \int_{x_L}^{+\infty} dx_R q_{-} (x_L,x_R) -   Q_{-}( x_L)  \right)  \right]
\label{c2.75}
\end{eqnarray}
can be understood as follows :
 the first line contains the constraints already present at the Level 2.5 of Eq. \ref{ld2.5rhoq},
while the two last lines contain the definitions of $Q_{\pm}(.)$ in terms of $q_{\pm}(.,.)$ discussed in Eqs \ref{jumpflowsrecover}.
The rate function corresponding to the alternate Markov chain of Eq. \ref{alternatemarkov}
with the explicit kernels of Eqs \ref{emoins} and \ref{eplus}
reads
\begin{small}
\begin{eqnarray}
&& I_{2.75} [ j_+(.),  Q_{\pm}(.), q_{\pm} (.,.)  ]   
\label{ratei2.75}
 \\
&& = \int_{-\infty}^{+\infty} dx_L \int_{x_L}^{+\infty} dx_R
\left[ q_{+} (x_R,x_L)  \ln \left( \frac{ q_{+} (x_R,x_L) }{ W_+(x_R \vert x_L) Q_-(x_L)  }   \right)
+ q_{-} (x_L,x_R)   \ln \left( \frac{ q_{-} (x_L,x_R)  }{ W_-(x_L \vert x_R) Q_+(x_R) }   \right)  
 \right]
 \nonumber \\
&& = \int_{-\infty}^{+\infty} dx_L \int_{x_L}^{+\infty} dx_R
\left[ q_{+} (x_R,x_L)  \ln \left( \frac{ q_{+} (x_R,x_L) }{ \frac{\gamma_+(x_R) }{v_+(x_R)} \ e^{ - \int_{x_L}^{x_R} dx \frac{\gamma_+( x )}{v_+(x)}  } Q_-(x_L)  }   \right)
+ q_{-} (x_L,x_R)   \ln \left( \frac{ q_{-} (x_L,x_R)  }{ \frac{\gamma_-(x_L) }{v_-(x_L)} \ e^{ - \int_{x_L}^{x_R} dx \frac{\gamma_-( x )}{v_-(x)}  } Q_+(x_R) }   \right)  
 \right]
\nonumber
\end{eqnarray}
\end{small}
and vanishes for
the steady state values values
\begin{eqnarray}
q^{*}_{+} (x_R,x_L) && =W_+(x_R \vert x_L) Q^{*}_-(x_L)
= \frac{\gamma_+(x_R) }{v_+(x_R)} \ e^{\displaystyle - \int_{x_L}^{x_R} dx \frac{\gamma_+( x )}{v_+(x)}  }
\frac{\gamma_{-}( x_L) }{v_-(x_L)}   j_+^*( x_L) 
\nonumber \\
q^{*}_{-} (x_L,x_R) && =W_-(x_L \vert x_R) Q^{*}_+(x_R)
=\frac{\gamma_-(x_L) }{v_-(x_L)} \ e^{\displaystyle - \int_{x_L}^{x_R} dx \frac{\gamma_-( x )}{v_-(x)}  }
\frac{ \gamma_{+}( x_R) }{v_+(x_R) } j_+^*( x_R) 
\label{qqtyp}
\end{eqnarray}

The link between the Level 2.75 of Eq. \ref{ld2.75}
and the Level 2.5 of Eq. \ref{ld2.5rhoq} is explained in Appendix \ref{app_contraction2.75to2.5}.


\section{ Conclusions}

\label{sec_conclusion}

In this paper, we have considered the one-dimensional run-and-tumble process on the infinite line, 
when the space-dependence of the two velocities $v_{\pm}(x)$ 
and/or of the two switching rates $\gamma_{\pm}(x)$ produces a localized non-equilibrium steady state.
The goal was to analyze its large deviations properties at various levels.

As explained in the Introduction, for non-equilibrium steady states of Markov processes,
one should start with the Level 2.5 that represents
 the lowest level where one can write the explicit rate function and the corresponding constraints.
For the one-dimensional run-and-tumble process, we have explained that the Level 2.5 characterizes
the joint probability of the two empirical densities $ \rho_{\pm}( x)  $, 
of the two empirical spatial currents $ j_{\pm}( x) $ and of the two empirical switching flows $ Q_{\pm}( x) $,
that are related via many constitutive constraints.
We have then described how the Level 2 for the empirical densities $ \rho_{\pm}( x)  $ alone can be obtained this Level 2.5 via contraction.
More generally, we have explained how the Level 2.5 can be contracted 
to obtain the scaled cumulant generating function of any time-additive observable.

We have then analyzed the large deviations at Level 2.75 for
the joint probability of the empirical intervals between consecutive switching events
via the introduction of the alternate Markov chain that governs the series of all the switching events of a long trajectory.  We have explained why this Level 2.75 contains more information that the Level 2.5 that can be recovered via contraction (see Appendix \ref{app_contraction2.75to2.5}). 
As explained in detail in the recent preprint \cite{c_jumpdriftdiff},
this property can be extended to other types of jump-drift processes,
 where the motion between jumps is also deterministic, 
 while the analysis is different for jump-diffusion processes, 
 where the motion between jumps is then stochastic.

As a final remark, let us mention that some large deviations properties of other active matter models
are discussed in the recent preprint \cite{jack_active}.


\appendix


\section{ Examples of run-and-tumble processes with localized steady states  }

\label{app_examples}

In this Appendix related to section \ref{sec_steady}, 
we describe some simple examples of localized non-equilibrium steady states
of one-dimensional run-and-tumble processes, 
where only the switching rates or only the velocities are space-dependent.

\subsection{ Steady states produced by 
uniform velocities $v_{\pm}(x)= v $ and space-dependent switching rates $\gamma_{\pm}(x)$  }

When the two velocities are uniform and equal $v_{\pm}(x)= v $, 
 the steady-state of Eqs \ref{pfromjp} and \ref{soljpst}
\begin{eqnarray}
P_{\pm}^*( x)  = \frac{ j^*_+(x) }{   v } = 
\frac{  j^*_+(0)}{v}  \ e^{\displaystyle \frac{1}{v} \int_0^x dy \left[  \gamma_-(y)    -  \gamma_+(y)   \right]}
\label{pvuni}
\end{eqnarray}
will be normalizable if the space-dependent switching rates $\gamma_{\pm}(x) $
ensure the convergence of the integral of Eq. \ref{stcv}
\begin{eqnarray}
\int_{-\infty}^{+\infty} dx 
 e^{\displaystyle  \frac{1}{v} \int_0^x dy \left[  \gamma_-(y)    -  \gamma_+(y)   \right]}
 < +\infty
\label{stcvgamma}
\end{eqnarray}

Among the various cases discussed in \cite{bressloff2021}, let us mention :

(i) an example with a smooth variation of the switching rates $\gamma_{\pm}(x)$ over the characteristic length $\xi$
between the minimum value $\gamma_0$ and the maximum value $\gamma>\gamma_0$
\begin{eqnarray}
  \gamma^{smooth}_{\pm}(x) = \frac{\gamma+\gamma_0}{2} \pm \frac{\gamma-\gamma_0}{2}   \tanh \frac{x}{\xi} 
\label{smooth}
\end{eqnarray}
with the corresponding localized steady state given by Eq. \ref{pvuni} 
\begin{eqnarray}
P_{\pm}^*( x) =
\frac{  j^*_+(0)}{v}  \ e^{\displaystyle - \frac{ ( \gamma-\gamma_0)}{v} \int_0^x dy \tanh \frac{y}{\xi}}
= \frac{  j^*_+(0)}{v}  \ e^{ \displaystyle  -  \frac{  ( \gamma-\gamma_0) \xi }{v}  \ln \left( \cosh \frac{x}{\xi} \right)}
 =  \frac{ j^*_+(0)  }{ v \left[ \cosh \frac{x}{\xi} \right]^{ \frac{  ( \gamma-\gamma_0) \xi }{v} } }
\label{pvunismooth}
\end{eqnarray}

(ii) an example with a step variation of the switching rates $\gamma_{\pm}(x)$ at the origin $x=0$ 
(corresponding to the limit $\xi \to +\infty$ of Eq. \ref{smooth})
\begin{eqnarray}
 \gamma^{step}_{\pm}(x) = \frac{\gamma+\gamma_0}{2} \pm \frac{\gamma-\gamma_0}{2}  {\rm sgn} (x)
\label{step}
\end{eqnarray}
or equivalently in terms of the Heaviside function $\theta(x)=\frac{1+{\rm sgn} (x)}{2}$
\begin{eqnarray}
 \gamma^{step}_{+}(x) = \gamma_0 +(\gamma-\gamma_0)  \theta (x)
 \nonumber \\
  \gamma^{step}_{-}(x) = \gamma - (\gamma-\gamma_0) \theta (x)
\label{steph}
\end{eqnarray}

The corresponding steady state of Eq. \ref{pvuni} is then given by the simple symmetric exponential form
\begin{eqnarray}
P_{\pm}^*( x)  =
\frac{  j^*_+(0)}{v}  \ e^{\displaystyle - \frac{ ( \gamma-\gamma_0) }{v} \int_0^x dy \  {\rm sgn} (y) }
= \frac{  j^*_+(0)}{v}  \ e^{\displaystyle - \frac{  ( \gamma-\gamma_0)}{v} \vert x \vert }
\label{pvunistep}
\end{eqnarray}

 
\subsection{ Steady states produced by 
space-dependent velocities $v_{\pm}(x)$ and uniform switching rates $\gamma_{\pm}(x)= \gamma_{\pm}$  }

When the two switching rates $\gamma_{\pm}(x)= \gamma_{\pm}$
do not depend on the position  $x$, 
the dynamics for the internal variable $\sigma=\pm 1$ 
is of course closed : the two probabilities
\begin{eqnarray}
P_{\sigma}( t)     && = \int_{-\infty}^{+\infty} dx   P_{\sigma}( x,t)
\label{psigmaalone}
\end{eqnarray}
then correspond to the so-called telegraph process 
\begin{eqnarray}
 \partial_t P_+( t)     && =   -   \gamma_+  P_+( t)+ \gamma_-  P_-( t)
\nonumber \\
 \partial_t P_-( t)    && =      \gamma_+  P_+( t) -   \gamma_-  P_-( t)
\label{telegraph}
\end{eqnarray}
and converge towards the following steady state for the internal variable $\sigma=\pm$
\begin{eqnarray}
 P_+^*    && =  \frac{ \gamma_-  }{ \gamma_+ + \gamma_-}
\nonumber \\
P_-^*    && =  \frac{ \gamma_+  }{ \gamma_+ + \gamma_-}
\label{telegraphst}
\end{eqnarray}
However, the dynamics for the position $x$ will have a localized steady-state
only if the two space-dependent velocities $v_{\pm}(x)$
satisfy the condition of Eq. \ref{stcv}
\begin{eqnarray}
 \int_{-\infty}^{+\infty} dx  \left[  \frac{ 1}{v_+( x)} + \frac{ 1   }{v_-(x) }  \right]
 e^{\displaystyle  \int_0^x dy \left[ \frac{ \gamma_-    }{v_-(y) } - \frac{ \gamma_+ }{v_+( y)}  \right]} < +\infty
\label{stcvvelo}
\end{eqnarray}

An interesting example is the exponential functional of the telegraph process $\sigma(t)$ 
studied previously in the context of anomalous diffusion in random media \cite{c_diffcorre}
\begin{eqnarray}
x(t)  = \int_0^t dt' e^{ \int_{t'}^t dt'' \sigma(t'') }
\label{kestenint}
\end{eqnarray}
while analogous exponential functionals of Brownian motion have been also much studied \cite{c_flux,c_hyperbolic,c_yor,yor}.

The dynamics of $x(t)$ defined by Eq. \ref{kestenint}
corresponds to the multiplicative stochastic process
\begin{eqnarray}
\dot x(t)  = 1 +  \sigma(t) x(t)
\label{kestenderi}
\end{eqnarray}
described by Eqs  \ref{run} when the two velocities display 
the following linear dependences with respect to the position $x$
\begin{eqnarray}
v_+(x) = x +1
\nonumber \\
v_-(x) = x-1
\label{forcelin}
\end{eqnarray}
While $v_+(x) $ is positive for any position $x\geq 0$,
the velocity $v_{-}(x) $ is positive only for $x \geq 1$.
As a consequence, the steady state $P_{\pm}^*(  x)  $ has for support $x \in ]1, +\infty[$
(while the interval $0 \leq x< 1$ will be visited only in the transient regime starting from the initial condition $x(t=0)=0$).

The solution of Eq. \ref{runstst} needs to be adapted to the support $x \in ]1, +\infty[$
in terms of the new integration constant $K$
\begin{eqnarray}
 j^*_+(x) =   K e^{\displaystyle  e^{   \gamma_-   \ln(x-1) - \gamma_+ \ln(x+1) }} 
 = K \frac{(x-1)^{\gamma_-}}{(x+1)^{\gamma_+} }
\label{soljpstkesten}
\end{eqnarray}
The corresponding steady state $ P^*_{\pm}( x )$ of Eq. \ref{pfromjp}
\begin{eqnarray}
P_+^*( x) && = \frac{ j^*_+(x) }{   x+1 }   = K \frac{(x-1)^{\gamma_-}}{(x+1)^{\gamma_++1} }
\nonumber \\
P_-^*( x) && = \frac{ j^*_+(x) }{   x-1 }   = K \frac{(x-1)^{\gamma_--1}}{(x+1)^{\gamma_+} }
\label{pfromjpk}
\end{eqnarray}
is always normalizable at the boundary $x \to 1^+$,
while their common power-law behavior for large $x$
\begin{eqnarray}
 P^*_{\pm}( x)   \opsimeq_{x \to +\infty}     \frac{  K   }
 { x^{ \gamma_+ - \gamma_- +1}    }
\label{stsolukestenasymp}
\end{eqnarray}
 is normalizable only if the two switching rates satisfy
\begin{eqnarray}
  \gamma_+> \gamma_-
\label{solum}
\end{eqnarray}
Then one can compute the normalizations of Eqs \ref{pfromjpk} in terms of the Gamma-function $\Gamma(.)$
\begin{eqnarray}
\int_1^{+\infty} dx P_+^*( x) && =  K \int_1^{+\infty} dx \frac{(x-1)^{\gamma_-}}{(x+1)^{\gamma_++1} }
= K  \frac{\Gamma( \gamma_+ - \gamma_-) \Gamma(  \gamma_-)}{2^{\gamma_+ - \gamma_-} \Gamma( \gamma_+)} \ \frac{\gamma_-}{\gamma_+}
\nonumber \\
\int_1^{+\infty} dx P_-^*( x) && =    K \int_1^{+\infty} dx\frac{(x-1)^{\gamma_--1}}{(x+1)^{\gamma_+} }
=K  \frac{\Gamma( \gamma_+ - \gamma_-) \Gamma(  \gamma_-)}{2^{\gamma_+ - \gamma_-} \Gamma( \gamma_+)}
\label{pfromjpknorma}
\end{eqnarray}
The compatibility with Eqs \ref{telegraphst} yields that the normalization constant $K$ reads
\begin{eqnarray}
K  = 
\ \frac{ \gamma_+ \Gamma( \gamma_+) 2^{\gamma_+ - \gamma_-} }
{ ( \gamma_+ + \gamma_- ) \Gamma(  \gamma_-) \Gamma( \gamma_+ - \gamma_-) }
\label{knormatelegraph}
\end{eqnarray}


\section{ Inference interpretation of the Level 2.5 }

\label{app_alternative}

In this Appendix related to section \ref{sec_2.5}, we describe 
another interpretation of the large deviations at Level 2.5 of Eq. \ref{ld2.5rhoq}
via the inverse problem of inference \cite{c_inference} :
from the data of a long dynamical trajectory, 
one computes the empirical time-averaged observables described above,
and one infers the best steady current $\hat j_+^*(x) $
and the best corresponding switching rates $\hat \gamma_{\pm}(x)$ of the model as follows.

(i) the best inferred steady current $\hat j_+^*(x) $ is simply the measured empirical current $j_+(x)$ 
\begin{eqnarray}
\hat j_+^* (x) \equiv j_+(x)
\label{hatjp}
\end{eqnarray}

(ii) the best inferred switching rates $\hat \gamma_{\pm}(x)$  are the rates that would make 
the switching flows typical with respect to the empirical densities or equivalently with respect to the empirical current $ j_+(x)$
\begin{eqnarray}
\hat \gamma_{\pm}(x) \equiv  \frac{  Q_{\pm}(x)  }{   \rho_{\pm}(x) } = \frac{  Q_{\pm}(x)  }{  \frac{j_+(x)}{v_{\pm}(x)} }
\label{hatgamma}
\end{eqnarray}

Via this change of variables from $[j_+(x),Q_{\pm}(x)]$ towards $[\hat j_+^* (x), \hat \gamma_{\pm}(x) ]$, 
the large deviations at Level 2.5 of Eq. \ref{ld2.5rhoq} translates into 
the joint probability to infer the two switching rates $\hat \gamma_{\pm}(x)$
and the corresponding steady current $\hat j_+^* (x) $ that they produce together
\begin{eqnarray}
 P^{Infer}_T[ \hat j_+^* (x), \hat \gamma_{\pm}(x) ]   \opsimeq_{T \to +\infty}  
C_{Infer} [ \hat j_+^* (x), \hat \gamma_{\pm}(x) ] 
e^{- \displaystyle T I_{Infer} [ \hat j_+^* (x), \hat \gamma_{\pm}(x) ] }
\label{ldinfer}
\end{eqnarray}
The rate function translated from Eq. \ref{rate2.5rhoq} reads
\begin{eqnarray}
I_{Infer} [ \hat j_+^* (x), \hat \gamma_{\pm}(x) ] && = 
  \int_{-\infty}^{+\infty}d  x \frac{\hat j_+^* (x)}{v_{+}(x)}
\left(  \hat \gamma_{+}(x)    
\ln \left( \frac{  \hat \gamma_{+}(x)   }{    \gamma_+(x)}  \right) 
 -   \hat \gamma_{+}(x)   +   \gamma_+(x)    \right)
\nonumber \\
&& +
  \int_{-\infty}^{+\infty}d  x \frac{\hat j_+^* (x)}{v_{-}(x)}  
\left(  \hat \gamma_{-}(x) 
  \ln \left( \frac{  \hat \gamma_{-}(x)   }{    \gamma_-(x) }  \right) 
 -    \hat \gamma_{-}(x)  + \gamma_-(x)   \right)  
\label{rate2.5rhoqinfer}
\end{eqnarray}
The constraints translated from Eq. \ref{ld2.5rhoq}
\begin{eqnarray}
C_{Infer} [ \hat j_+^* (x), \hat \gamma_{\pm}(x) ] =  
\delta \left( \int_{-\infty}^{+\infty} dx \left[  \frac{1}{v_+(x) }   +  \frac{1}{v_-(x) } \right] \hat j^*_+(x)   - 1  \right)
 \left[ \prod_{ x } \delta \left(  \frac{d \hat j_+^*(x) }{dx}  
  + \hat j_+^* (x) \left[   \frac{\hat \gamma_{+}(x) }{v_{+}(x)} 
  -  \frac{\hat \gamma_{-}(x) }{v_{-}(x)} \right]    \right)  \right] 
\label{cinger}
\end{eqnarray}
means that the current $ j_+^*(x)$ is simply the steady current associated to the inferred rates $\hat \gamma_{\pm}(x) $ : indeed, the second constraint is the analog of Eq. \ref{eqjst}, while the first constraint
is the analog of Eq. \ref{stnorma}. As a consequence, 
the constraints can be fully solved to rewrite the current $ j_+^*(x)$
in terms of the inferred switching rates $\hat \gamma_{\pm}(x) $ (analog of Eqs \ref{soljpst} and \ref{stnorma})
\begin{eqnarray}
\hat j^*_+(x) =  \frac{ \ e^{\displaystyle  \int_0^x dy \left[ \frac{ \hat \gamma_-(y)    }{v_-(y) } - \frac{\hat \gamma_+(y) }{v_+( y)}  \right]} } { \displaystyle \int_{-\infty}^{+\infty} dz \left[  \frac{ 1 }{   v_+( z ) } + \frac{ 1 }{   v_-( z ) }  \right] 
 e^{\displaystyle  \int_0^z dy \left[ \frac{ \hat \gamma_-(y)    }{v_-(y) } - \frac{ \hat \gamma_+(y) }{v_+( y)}  \right]}} 
\label{soljpstinfer}
\end{eqnarray}
Plugging this expression into the rate function of Eq. \ref{rate2.5rhoqinfer}
\begin{eqnarray}
&& I_{Infer} [  \hat \gamma_{\pm}(x) ]  = 
\label{rate2.5rhoqinfergamma}  \\
  &&   
  \frac{ \displaystyle  \int_{-\infty}^{+\infty}d  x 
 \left[ \frac{  \hat \gamma_{+}(x)    
\ln \left( \frac{  \hat \gamma_{+}(x)   }{    \gamma_+(x)}  \right) 
 -   \hat \gamma_{+}(x)   +   \gamma_+(x)    }{v_{+}(x)}
+ \frac{  \hat \gamma_{-}(x) 
  \ln \left( \frac{  \hat \gamma_{-}(x)   }{    \gamma_-(x) }  \right) 
 -    \hat \gamma_{-}(x)  + \gamma_-(x)   }{v_{-}(x)}   \right] \ e^{\displaystyle  \int_0^x dy \left[ \frac{ \hat \gamma_-(y)    }{v_-(y) } - \frac{\hat \gamma_+(y) }{v_+( y)}  \right]} } 
 { \displaystyle \int_{-\infty}^{+\infty} dz \left[  \frac{ 1 }{   v_+( z ) } + \frac{ 1 }{   v_-( z ) }  \right] 
 e^{\displaystyle  \int_0^z dy \left[ \frac{ \hat \gamma_-(y)    }{v_-(y) } - \frac{ \hat \gamma_+(y) }{v_+( y)}  \right]}} 
\nonumber
\end{eqnarray}
one obtains that the joint probability to infer the two switching rates $\hat \gamma_{\pm}(x) $
instead of the true values $ \gamma_{\pm}(x) $
reduces to 
\begin{eqnarray}
 P^{Infer}_T[  \hat \gamma_{\pm}(x) ]   \opsimeq_{T \to +\infty}  
e^{- \displaystyle T I_{Infer} [  \hat \gamma_{\pm}(x) ] }
\label{ldinfergammaonly}
\end{eqnarray}


\section{ Explicit contraction from the Level 2.5 to the Level 2.25}

\label{app_contractionfrom2.5to2.25}

In this Appendix related to section \ref{sec_2}, we describe how the Level 2.5 of Eq. \ref{ld2.5rhoq}
produces the Level 2.25 of Eq. \ref{ld2.5diffwithoutA} via contraction.


\subsection{ Large deviations at level 2.5 in terms of the switching activity $A(x)$ 
and of the switching current $J( x)$ }

In the Level 2.5 of Eq. \ref{ld2.5rhoq},
the two switching flows $Q_{\pm}(x) $
can be replaced by their symmetric and antisymmetric parts called the activity and the current
\begin{eqnarray}
A( x)  && \equiv Q_{+}( x)  + Q_{-}( x)  
\nonumber \\
J( x)  && \equiv Q_{+}( x)  - Q_{-}( x)  
\label{currentactivity}
\end{eqnarray}
i.e. reciprocally
\begin{eqnarray}
Q_{+}( x) && = \frac{ A( x) + J( x)}{2}
\nonumber \\
Q_{-}( x) && = \frac{ A( x) - J( x)}{2}
\label{currentactivityinv}
\end{eqnarray}
The large deviation form of Eq. \ref{ld2.5rhoq}
translates into
\begin{eqnarray}
 P_T[  j_+(.), J(.),A(.)]   \opsimeq_{T \to +\infty} 
&&  \delta \left( \int_{-\infty}^{+\infty}  dx \left[  \frac{1}{v_+(x) }   +  \frac{1}{v_-(x) } \right] j_+(x)   - 1  \right)
 \left[ \prod_{ x } \delta \left( J( x)  + \frac{d j_+(x) }{dx}    \right)  \right]  
\nonumber \\ && e^{- \displaystyle T  I_{2.5}[ j_+(.),  J(.),A(.) ]    }
\label{ld2.5rhoaj}
\end{eqnarray}
with the rate function translated from Eq. \ref{rate2.5rhoq}
\begin{eqnarray}
&&  I_{2.5}[ j_+,  J(.),A(.) ]  
 = 
 \int_{-\infty}^{+\infty}d  x
\left[   \left( \frac{ A( x) +  J( x)}{2} \right) 
  \ln \left( \frac{    A( x) +  J( x)  }{  2   \frac{\gamma_+(x)}{v_+(x) }  j_+(x)     }  \right) 
    \right]  
 \nonumber \\
&& + 
  \int_{-\infty}^{+\infty}d  x
\left[   \left( \frac{ A( x) - J( x)}{2} \right)   \ln \left( \frac{     A( x) - J( x)   }
{   2    \frac{\gamma_-(x)}{v_-(x) }  j_+(x)     }  \right) 
 -  A( x)    +  \frac{\gamma_+(x)}{v_+(x) }  j_+(x)   +   \frac{\gamma_-(x)}{v_-(x) }  j_+(x)   \right]  
\label{rate2.5rhoaj}
\end{eqnarray}


\subsection{ Explicit contraction over the switching activity $A(x)$  }

Since the switching activity $A(x)$ does not appear in the constraints of Eq. \ref{ld2.5rhoaj},
one can optimize the rate function of Eq. \ref{rate2.5rhoaj}
over the activity as in many other Markov jump processes \cite{maes_canonical,c_ring,c_interactions,c_detailed}
\begin{eqnarray}
0= && \frac{ \partial  I_{2.5}[ \rho_{\pm}(.),  J(.),A(.) ]  }{\partial A(x) }
 = \frac{1}{2}  
  \ln \left( \frac{    A^2( x) -  J^2( x)  }{  4    \frac{\gamma_+(x)\gamma_-(x)}{v_+(x)v_-(x)}   j_+^2( x)   }  \right) 
\label{deria}
\end{eqnarray}
in order to obtain the optimal activity as a function of the spatial current $j_+(x) $ and of the switching current $J(x) $
\begin{eqnarray}
   A^{opt}[ j_+(.), J(.) ] 
   = \sqrt{   J^2( x)  + 4  \frac{\gamma_+(x)\gamma_-(x)}{v_+(x)v_-(x)}   j_+^2( x)     }
\label{aopt}
\end{eqnarray}
The corresponding rate function obtained via this explicit contraction 
\begin{eqnarray}
 I_{2.25} [ j_+(.),  J(.)] &&= I_{2.5} [ j_+(.),  J(.), A^{opt}[ j_+(.), J(.) ] ]
\label{rate2.25deri}
\end{eqnarray}
is given in Eq. \ref{rate2.25} of the main text.


\section{ Level 2.5 and Level 2 when the two switching rates $\gamma_{\pm}(x) $ have disjoint supports}

\label{app_disjoint}

In this Appendix, we describe how the Level 2.5 of section \ref{sec_2.5}
and the Level 2 of section \ref{sec_2}
can be simplified when the two switching rates $\gamma_{\pm}(x) $ have disjoint supports.
To be concrete, let us consider the case where the switching rates vanish when the particle is going back towards its home 
at the origin $x=0$,
i.e. $\gamma_{+}(x) $ vanishes in the whole region $x \in ]-\infty,0[$ and $\gamma_{-}(x) $ vanishes in the whole region $x \in ]0,+\infty[$
\begin{eqnarray}
 \gamma_{+}(x<0) = 0
 \nonumber \\
  \gamma_{-}(x>0) = 0
\label{disjoint}
\end{eqnarray}
The condition of Eq. \ref{stcv} for the existence of a localized steady state becomes
\begin{eqnarray}
\int_{-\infty}^{0} dx \left[  \frac{ 1 }{   v_+( x ) } + \frac{ 1 }{   v_-( x ) }  \right] 
 e^{\displaystyle  - \int_x^0 dy  \frac{ \gamma_-(y)    }{v_-(y) } }
 +
 \int_{0}^{+\infty} dx \left[  \frac{ 1 }{   v_+( x ) } + \frac{ 1 }{   v_-( x ) }  \right] 
 e^{\displaystyle  - \int_0^x dy  \frac{ \gamma_+(y) }{v_+( y)}  }
 < +\infty
\label{stcvfactor}
\end{eqnarray}

\subsection{ Simplifications for the Level 2.5 }

Since the empirical switching flow $Q_+(x)$ has for support $x \in ]0,+\infty[$
and the empirical switching flow $Q_-(x)$ has for support $x \in ]-\infty,0[$,
the integral version of the stationary constraint of Eq. \ref{jpintegral} becomes
\begin{eqnarray}
  j_+( x) && = \int_x^{+\infty} dy Q_{+}( y)   \ \ {\rm for } \ x \geq 0
 \nonumber \\ 
  j_+( x)  &&  =   \int_{-\infty}^x dy  Q_{-}( y)  \ \ {\rm for } \ x \leq 0
\label{jpintegralfactor}
\end{eqnarray}
while the consistency of the two expressions at the origin $x=0$ ensures
that the total density of switching events out of the state $+$ and out of the state $-$ are equal (Eq. \ref{vanishing}).

The large deviations at Level 2.5 of Eqs \ref{ld2.5rhoq}
can be then factorized into the two regions $x>0$ and $x<0$ 
except for the junction concerning the current $ j_+( x=0) $ at the origin and the global normalization
\begin{eqnarray}
&& P_T[  j_+(.) , Q_{\pm}(.) ]   \opsimeq_{T \to +\infty} 
  \delta \left( \int_{-\infty}^{+\infty} dx \left[  \frac{1}{v_+(x) }   +  \frac{1}{v_-(x) } \right] j_+(x)   - 1  \right)
 \nonumber \\
 && \left[ \prod_{ x\geq 0 } \delta \left(   j_+( x) - \int_x^{+\infty} dy Q_{+}( y)   \right)  \right] 
 e^{ \displaystyle - T   \int_{0}^{+\infty}d  x
\left[  Q_{+}( x)    \ln \left( \frac{  Q_{+}( x)   }{     \frac{\gamma_+(x)}{v_+(x) }  j_+(x) }  \right) 
 -   Q_{+}( x)   +    \frac{\gamma_+(x)}{v_+(x) }  j_+(x)    \right]     } 
   \nonumber \\
&&   \left[ \prod_{ x\leq 0 } \delta \left(    j_+( x)  -\int_{-\infty}^x dy  Q_{-}( y)   \right)  \right] 
e^{ \displaystyle - T  \int_{-\infty}^{0}d  x
\left[  Q_{-}( x)   \ln \left( \frac{  Q_{-}( x)   }{    \frac{\gamma_-(x)}{v_-(x) }  j_+(x) }  \right) 
 -   Q_{-}( x)   +   \frac{\gamma_-(x)}{v_-(x) }  j_+(x)  \right]     }
\label{ld2.5rhoqfactor}
\end{eqnarray}

\subsection{ Simplifications for the Level 2 }

If one eliminates the empirical switching flows via
\begin{eqnarray}
Q_+(x) && = -  j_+'( x)    \ \ {\rm for } \ x \geq 0
 \nonumber \\ 
 Q_-(x)  &&  =  j_+'( x)    \ \ {\rm for } \ x \leq 0
\label{jpintegralelim}
\end{eqnarray}
one obtains that the probability for the empirical spatial current $j_+(x)$ alone 
follows the large deviation form of Eq. \ref{ld2rhoalone}
with the rate function 
\begin{footnotesize}
\begin{eqnarray}
&&  I_2 [ j_+(.)] 
   = 
 \int_{0}^{+\infty}d  x
\left[  -  j_+'( x)     \ln \left( \frac{  (-  j_+'( x) )   }{     \frac{\gamma_+(x)}{v_+(x) }  j_+(x) }  \right) 
 + j_+'( x)    +    \frac{\gamma_+(x)}{v_+(x) }  j_+(x)    \right]   
 + \int_{-\infty}^{0}d  x
\left[  j_+'( x)   \ln \left( \frac{  j_+'( x)  }{    \frac{\gamma_-(x)}{v_-(x) }  j_+(x) }  \right) 
 -   j_+'( x)  +   \frac{\gamma_-(x)}{v_-(x) }  j_+(x)  \right]    
 \nonumber \\ 
&&
= \int_{0}^{+\infty}d  x
\left[  -  j_+'( x)     \ln \left( \frac{  (-  j_+'( x) )   }{     \frac{\gamma_+(x)}{v_+(x) }  }  \right) 
 + \frac{d [ j_+( x) \ln j_+(x) ]}{dx}    +    \frac{\gamma_+(x)}{v_+(x) }  j_+(x)    \right]   
 + \int_{-\infty}^{0}d  x
\left[  j_+'( x)   \ln \left( \frac{  j_+'( x)  }{    \frac{\gamma_-(x)}{v_-(x) }   }  \right) 
 -  \frac{d[ j_+( x) \ln j_+(x) ]}{dx}  +   \frac{\gamma_-(x)}{v_-(x) }  j_+(x)  \right]    
  \nonumber \\ 
&&
= - 2 j_+( 0) \ln j_+(0) 
+   \int_{0}^{+\infty}d  x
\left[  -  j_+'( x)     \ln \left( \frac{  (-  j_+'( x) )   }{     \frac{\gamma_+(x)}{v_+(x) }  }  \right) 
 +    \frac{\gamma_+(x)}{v_+(x) }  j_+(x)    \right]   
 + \int_{-\infty}^{0}d  x
\left[  j_+'( x)   \ln \left( \frac{  j_+'( x)  }{    \frac{\gamma_-(x)}{v_-(x) }   }  \right) 
  +   \frac{\gamma_-(x)}{v_-(x) }  j_+(x)  \right]   
    \nonumber \\ 
&&
=   \int_{0}^{+\infty}d  x
\left[  -  j_+'( x)     \ln \left( \frac{  (-  j_+'( x) )   }{     \frac{\gamma_+(x)}{v_+(x) } j_+( 0) }  \right) 
 +    \frac{\gamma_+(x)}{v_+(x) }  j_+(x)    \right]   
 + \int_{-\infty}^{0}d  x
\left[  j_+'( x)   \ln \left( \frac{  j_+'( x)  }{    \frac{\gamma_-(x)}{v_-(x) } j_+( 0)  }  \right) 
  +   \frac{\gamma_-(x)}{v_-(x) }  j_+(x)  \right]       
  \label{i2rhoalonefactor}
\end{eqnarray}
\end{footnotesize}

\subsection{ Large deviations for the switching flows $Q_{\pm}(x)$ alone }

If one wishes instead to eliminate the empirical spatial current $j_+(x)$ in terms of the two 
the empirical switching flows $Q_{\pm}(x)$ via Eqs \ref{jpintegralfactor},
the normalization constraint becomes
\begin{eqnarray}
 1 && = \int_{-\infty}^{+\infty} dx \left[  \frac{1}{v_+(x) }   +  \frac{1}{v_-(x) } \right] j_+(x) 
\nonumber \\
&& = \int_{-\infty}^{0} dx \left[  \frac{1}{v_+(x) }   +  \frac{1}{v_-(x) } \right]  \int_{-\infty}^x dy  Q_{-}( y) 
 + \int_{0}^{+\infty} dx \left[  \frac{1}{v_+(x) }   +  \frac{1}{v_-(x) } \right]  \int_x^{+\infty} dy Q_{+}( y)
 \nonumber \\
&& =  \int_{-\infty}^0 dy  Q_{-}( y)  \int_{y}^{0} dx \left[  \frac{1}{v_+(x) }   +  \frac{1}{v_-(x) } \right] 
 + \int_0^{+\infty} dy Q_{+}( y) \int_{0}^{y} dx \left[  \frac{1}{v_+(x) }   +  \frac{1}{v_-(x) } \right]  
 \label{normaforQ}
\end{eqnarray}
while the total density $n$ of the switching events of each kind is given by
\begin{eqnarray}
n \equiv  \int_0^{+\infty} dy Q_{+}( y)  =   \int_{-\infty}^0 dy  Q_{-}( y)  =  j_+( 0)  
\label{densitynfactor}
\end{eqnarray}
and the rate function of Eq. \ref{i2rhoalonefactor} translates into
\begin{footnotesize}
\begin{eqnarray}
&&  I_2 [ n,Q_{\pm}(.)] 
\nonumber \\ &&
=   \int_{0}^{+\infty}d  x Q_+( x)     \ln \left( \frac{ Q_+( x)    }{     \frac{\gamma_+(x)}{v_+(x) } n }  \right) 
 +   \int_{0}^{+\infty}d  x  \frac{\gamma_+(x)}{v_+(x) } \int_x^{+\infty} dy Q_{+}( y)    
+ \int_{-\infty}^{0}d  x
 Q_-( x)   \ln \left( \frac{ Q_-( x)  }{    \frac{\gamma_-(x)}{v_-(x) } n  }  \right) 
  +  \int_{-\infty}^{0}d  x \frac{\gamma_-(x)}{v_-(x) }  \int_{-\infty}^x dy  Q_{-}( y)    
\nonumber \\
&&  =   \int_{0}^{+\infty}d  x Q_+( x)     \ln \left( \frac{ Q_+( x)    }{     \frac{\gamma_+(x)}{v_+(x) } n }  \right) 
 +  \int_0^{+\infty} dy Q_{+}( y)   \int_{0}^{y}d  x  \frac{\gamma_+(x)}{v_+(x) }   
 + \int_{-\infty}^{0}d  x
 Q_-( x)   \ln \left( \frac{ Q_-( x)  }{    \frac{\gamma_-(x)}{v_-(x) } n  }  \right) 
  + \int_{-\infty}^0 dy  Q_{-}( y)  \int_{y}^{0}d  x \frac{\gamma_-(x)}{v_-(x) }   
  \nonumber \\
&&  =   \int_{0}^{+\infty}d  x Q_+( x)     \ln \left( \frac{ Q_+( x)    }{     \frac{\gamma_+(x)}{v_+(x) } 
e^{-  \int_{0}^{x}d  y  \frac{\gamma_+(y)}{v_+(y) } } 
n }  \right) 
 + \int_{-\infty}^{0}d  x
 Q_-( x)   \ln \left( \frac{ Q_-( x)  }{    \frac{\gamma_-(x)}{v_-(x) }
 e^{- \int_{x}^{0}d  y \frac{\gamma_-(y)}{v_-(y) }}
  n  }  \right)  
  \label{i2rhoalonefactorq}
\end{eqnarray}
\end{footnotesize}
Putting everything together, 
one obtains that the joint distribution of the empirical switching flows $Q_{\pm}(x)$
and of the total density $n$ of switching events of each kind
follows the large deviation form
\begin{eqnarray}
&& P_T[ n , Q_{\pm}(.) ]   \opsimeq_{T \to +\infty} 
 \delta \left(  n -  \int_0^{+\infty} dx Q_{+}( x)  \right)  \delta \left(  n -  \int_{-\infty}^0 dx  Q_{-}( x) \right)
\nonumber \\
&&  \delta \left(  \int_{-\infty}^0 dx  Q_{-}( x)  \int_{x}^{0} dy \left[  \frac{1}{v_+(y) }   +  \frac{1}{v_-(y) } \right] 
 + \int_0^{+\infty} dx Q_{+}( x) \int_{0}^{x} dy \left[  \frac{1}{v_+(y) }   +  \frac{1}{v_-(y) } \right]  
 - 1  \right)
 \nonumber \\
 && 
 e^{ \displaystyle - T   \int_{0}^{+\infty}d  x Q_+( x)     \ln \left( \frac{ Q_+( x)    }
 {     \frac{\gamma_+(x)}{v_+(x) } e^{-  \int_{0}^{x}d  y  \frac{\gamma_+(y)}{v_+(y) } }   n }  \right) 
 + \int_{-\infty}^{0}d  x
 Q_-( x)   \ln \left( \frac{ Q_-( x)  }{    \frac{\gamma_-(x)}{v_-(x) }
 e^{- \int_{x}^{0}d  y \frac{\gamma_-(y)}{v_-(y) }}
  n  }  \right)       }
\label{ld2.5rhoqfactorqalone}
\end{eqnarray}

\subsection{ Physical interpretation in terms of alternate excursions }

In Eq. \ref{ld2.5rhoqfactorqalone},
 one recognizes the standard form of rate function for semi-Markov processes 
\cite{fortelle_thesis,c_largedevdisorder,gaspard,maes_semi,zambotti,faggionato,c_reset}.
The physical meaning in terms of the alternate excursions on the right and on the left of the origin $x=0$
 can be understood as follows : their common density $n$ is given by the two constraints of the first line ;
 $Q_+(x)$ represents the 
empirical density of positive excursions ending at position $x$
and consuming the total time (for the forward travel from $0$ to $x>0$ at the velocity $v_+(.)$
and for the backward travel from $x$ to $0$ at the velocity $v_-(.)$)
\begin{eqnarray}
\tau_+(x)  = \int_{0}^{x} dy \left[  \frac{1}{v_+(y) }   +  \frac{1}{v_-(y) } \right] 
\label{tauplus}
\end{eqnarray}
while $Q_-(x)$ represents the 
empirical density of negative excursions ending at position $x$
and consuming the total time (for the travel from $0$ to $x<0$ at the velocity $v_-(.)$
and for the travel from $x$ to $0$ at the velocity $v_+(.)$)
\begin{eqnarray}
\tau_-(x) && = \int_{x}^{0} dy \left[  \frac{1}{v_+(y) }   +  \frac{1}{v_-(y) } \right] 
\label{taumoins}
\end{eqnarray}
so that the constraint of the second line corresponds to the normalization of the total time consumed during these excursions.

For a positive excursion, the 'true' probability to end at position $x$ appears in the rate function of Eq. \ref{ld2.5rhoqfactorqalone}
\begin{eqnarray}
{\cal P}^{exc}_+(x)  =\frac{\gamma_+(x)}{v_+(x) } e^{-  \int_{0}^{x}d  y  \frac{\gamma_+(y)}{v_+(y) } } 
\label{pexcplus}
\end{eqnarray}
with the normalization
\begin{eqnarray}
\int_0^{+\infty} dx {\cal P}^{exc}_+(x)  = 1
\label{pexcplusnorma}
\end{eqnarray}
Similarly for a negative excursion, the 'true' probability to end at position $x$ appears in the rate function of Eq. \ref{ld2.5rhoqfactorqalone}
\begin{eqnarray}
{\cal P}^{exc}_-(x)  =\frac{\gamma_-(x)}{v_-(x) } e^{-  \int_{x}^{0}d  y  \frac{\gamma_-(y)}{v_-(y) } } 
\label{pexcmoins}
\end{eqnarray}
with the normalization
\begin{eqnarray}
\int_{-\infty}^0 dx {\cal P}^{exc}_-(x)  = 1
\label{pexcmoinsnorma}
\end{eqnarray}

The similarity with the large deviations properties in some models of stochastic resetting \cite{c_reset}
can be explained as follows.
The present model of Eq. \ref{disjoint}
can actually be re-interpreted as some king of stochastic resetting where the reset events are not instantaneous but 
consume some return time towards the origin :
when a switching event occurs at $x>0$, one could say that there is a reset towards the origin $x=0$ in the negative state $\sigma=-$
that consumes the return time duration $\int_{0}^{x}   \frac{dy}{v_-(y) }   $ ;
similarly when a switching event occurs at $x<0$, one could say that there is a reset towards the origin $x=0$ in the positive state $\sigma=+$ that consumes the return time duration $\int_{x}^{0}   \frac{dy}{v_+(y) }   $.

This analysis of excursions between returns to the origin is of course very specific to the present model of Eq. \ref{disjoint}.
For the general run-and-tumble process,
the large deviations for the intervals between two consecutive switching events are analyzed in section \ref{sec_intervals}.


\section { Computation of the two first cumulants of any time-additive variable $A_T$}

\label{app_perturbativeoptimization}

In this Appendix related to section \ref{sec_additive}, we describe how the optimization procedure summarized in subsection \ref{subsec_optimization3steps}
can be implemented at the level of the perturbation theory in the parameter $k$ up to order $k^2$
in order to obtain the two first cumulants of the time-additive variable $A_T$ of Eq. \ref{additiveAdiff}.
(If one is interested into scaled higher cumulants, one needs to add higher orders $k^3,k^4,..$ into 
the perturbative framework described below).

Plugging the perturbative expansions of the scaled cumulant generating function
\begin{eqnarray}
\mu(k) && =   k \mu^{(1)} + k^2 \mu^{(2)}  + O(k^3)
  \label{muexp}
\end{eqnarray}
and of the Lagrange multiplier 
\begin{eqnarray}
\lambda_k(x) && =   k \lambda^{(1)}(x) + k^2 \lambda^{(2)}(x) + O(k^3)
  \label{lambdaexp}
\end{eqnarray}
 into Eq. \ref{lagrangederi3expli}
yields the following differential equations for $ \lambda^{(1)}(x) $ and $ \lambda^{(2)}(x) $ at order $k$ and $k^2$ respectively
 \begin{eqnarray}
 \frac{d \lambda^{(1)}(x)}{dx}  + \left[ \frac{ \gamma_-(x)    }{v_-(x) } - \frac{ \gamma_+(x) }{v_+( x)}  \right]\lambda^{(1)}(x)
&& =  \Upsilon_1(x)
\nonumber  \\
 \frac{d \lambda^{(2)}(x) }{dx} + \left[ \frac{ \gamma_-(x)    }{v_-(x) } - \frac{ \gamma_+(x) }{v_+( x)}  \right]\lambda^{(2)}(x)
  && =\Upsilon_2(x)
\label{lagrangederi3per1}  
  \end{eqnarray}
with the inhomogeneous terms 
 \begin{eqnarray}
\Upsilon_1(x)
&& \equiv  \left[ \frac{ 1   }{v_+(x) } + \frac{ 1 }{v_-( x)}  \right]  \mu^{(1)}
  - \frac{       \gamma_{+}( x) \beta_+(x)       }{v_+(x) } 
 - \frac{     \gamma_{-}( x) \beta_-(x)   }{v_-(x) }  -   \omega(x)     
\nonumber
 \\
\Upsilon_2(x)
&& \equiv  \left[ \frac{ 1   }{v_+(x) } + \frac{ 1 }{v_-( x)}  \right]  \mu^{(2)}
   -  \frac{  \gamma_{+}( x)    [\beta_+(x) -  \lambda^{(1)}(x)]^2      }{2 v_+(x) } 
 - \frac{  \gamma_{-}( x)   [\beta_-(x) + \lambda^{(1)}(x) ]^2   }{2 v_-(x) } 
\label{upsilon}
\end{eqnarray}
The solutions of Eqs \ref{lagrangederi3per1} that do not diverge exponentially at $x \to \pm \infty$ read 
 \begin{eqnarray}
 \lambda^{(1)}(x) && = e^{\displaystyle -  \int_0^x dz \left[ \frac{ \gamma_-(z)    }{v_-(z) } - \frac{ \gamma_+(z) }{v_+( z)}  \right]}
\int_{-\infty}^x dy   \Upsilon_1(y)e^{\displaystyle   \int_0^y dz \left[ \frac{ \gamma_-(z)    }{v_-(z) } - \frac{ \gamma_+(z) }{v_+( z)}  \right]}
\nonumber  \\
 \lambda^{(2)}(x) && = e^{\displaystyle -  \int_0^x dz \left[ \frac{ \gamma_-(z)    }{v_-(z) } - \frac{ \gamma_+(z) }{v_+( z)}  \right]}
\int_{-\infty}^x dy   \Upsilon_2(y)e^{\displaystyle   \int_0^y dz \left[ \frac{ \gamma_-(z)    }{v_-(z) } - \frac{ \gamma_+(z) }{v_+( z)}  \right]}
\label{lagrangederi3persolu}  
  \end{eqnarray}
  and require the vanishing of the following integrals involving the inhomogeneous terms of Eq. \ref{upsilon}
   \begin{eqnarray}
0 && = 
\int_{-\infty}^{+\infty} dx   \Upsilon_1(x)e^{\displaystyle   \int_0^x dz \left[ \frac{ \gamma_-(z)    }{v_-(z) } - \frac{ \gamma_+(z) }{v_+( z)}  \right]}
\label{conditionfullintegral} 
 \\
&& = \int_{-\infty}^{+\infty} dx  e^{\displaystyle   \int_0^x dz \left[ \frac{ \gamma_-(z)    }{v_-(z) } - \frac{ \gamma_+(z) }{v_+( z)}  \right]}
\left( \left[ \frac{ 1   }{v_+(x) } + \frac{ 1 }{v_-( x)}  \right]  \mu^{(1)}
  - \frac{       \gamma_{+}( x) \beta_+(x)       }{v_+(x) } 
 - \frac{     \gamma_{-}( x) \beta_-(x)   }{v_-(x) }  -   \omega(x)     
 \right)
\nonumber  \\
0 && = 
\int_{-\infty}^{+\infty} dx   \Upsilon_2(x)e^{\displaystyle   \int_0^x dz \left[ \frac{ \gamma_-(z)    }{v_-(z) } - \frac{ \gamma_+(z) }{v_+( z)}  \right]}
\nonumber \\
&&= 
\int_{-\infty}^{+\infty} dx   e^{\displaystyle   \int_0^x dz \left[ \frac{ \gamma_-(z)    }{v_-(z) } - \frac{ \gamma_+(z) }{v_+( z)}  \right]}
\left(  \left[ \frac{ 1   }{v_+(x) } + \frac{ 1 }{v_-( x)}  \right]  \mu^{(2)}
   -  \frac{  \gamma_{+}( x)    [\beta_+(x) -  \lambda^{(1)}(x)]^2      }{2 v_+(x) } 
 - \frac{  \gamma_{-}( x)   [\beta_-(x) + \lambda^{(1)}(x) ]^2   }{2 v_-(x) }\right)
\nonumber  
  \end{eqnarray}
  that will determine the values of $\mu^{(1)}$ and $\mu^{(2)}$ respectively.
 Using the explicit forms of the steady state $P_{\pm}^*(.)$ of Eqs \ref{pfromjp} and 
 of the steady state current $j_+^{*}(x) $ of Eq. \ref {soljpst},
 one obtains that the two first cumulants read
    \begin{eqnarray}
 \mu^{(1)} &&  
 = \int_{-\infty}^{+\infty} dx  \left( \frac{       \gamma_{+}( x) \beta_+(x)       }{v_+(x) } 
 + \frac{     \gamma_{-}( x) \beta_-(x)   }{v_-(x) }  +  \omega(x)      \right) j_+^{*}(x)
\nonumber  \\
 \mu^{(2)} 
&&= 
\int_{-\infty}^{+\infty} dx  \left(    \frac{  \gamma_{+}( x)    [\beta_+(x) -  \lambda^{(1)}(x)]^2      }{2 v_+(x) } 
 + \frac{  \gamma_{-}( x)   [\beta_-(x) + \lambda^{(1)}(x) ]^2   }{2 v_-(x) }\right) j_+^{*}(x)
\label{solumu12}
  \end{eqnarray}
  The first cumulant $  \mu^{(1)} $ depends only on the steady state as it should,
  while the scaled variance $  \mu^{(2)}$ involves in addition the first-order solution $\lambda^{(1)}(x)$ 
  of Eq. \ref{lagrangederi3persolu}.


\section { Link with the deformed generator method for additive observables }

\label{app_doob}

In this Appendix related to section \ref{sec_additive}, we describe the standard deformed generator method to analyze the large deviations of 
additive observables in order to explain the equivalence with the contraction from the Level 2.5
described in section \ref{sec_additive}.
 
\subsection { The scaled cumulant generating function $\mu(k)$ from the deformed generator method }
 
 The generating function of the additive observable $A_T$ of Eq. \ref{additiveAdiff}
 as parametrized by the six functions $(\alpha_{\pm}(x);\beta_{\pm}(x);\omega_{\pm}(x))$
 \begin{eqnarray}
    < e^{ \displaystyle T k A_T } >  
=    < e^{ \displaystyle  k \left[
\int_0^T dt  \left[ \alpha_{\sigma(t)}(  x(t)) + \dot x(t) \omega_{\sigma(t)}(  x(t))\right]
+   \sum_{t  \in [0,T]: \sigma(t^+) \ne \sigma(t^-) } \beta_{\sigma(t^-)}(x(t) )
\right] } >  \opsimeq_{T \to +\infty} e^{T \mu(k)}
\label{genekdiff}
\end{eqnarray}
can be analyzed via the following non-conserved deformed generator 
\begin{eqnarray}
 \frac{ \partial {\tilde P}_+( x,t)   }{\partial t}  &&   =  
 -    \left[  \partial_x  - k \omega_+(x) \right] \left[ v_+(x) {\tilde P}_+( x ,t)  \right]  
+  \left[ k \alpha_+ ( x ) -\gamma_+(x)  \right]    {\tilde P}_+( x,t)
 + \gamma_- (x) e^{k \beta_-(x) }{\tilde P}_-( x,t)
 \nonumber \\
  \frac{ \partial {\tilde P}_-( x,t)   }{\partial t}  &&   =  
       \left[ \partial_x - k \omega_-(x) \right] \left[ v_-(x) {\tilde P}_-( x,t )  \right]
    + \gamma_+(x) e^{k \beta_+(x) } {\tilde P}_+( x,t)
+  \left[ k \alpha_- ( x ) -\gamma_-(x)  \right]   {\tilde P}_-( x,t)
\label{tilted}
\end{eqnarray}

The scaled cumulant generating function $\mu(k)$ governing the asymptotic behavior of Eq. \ref{genekdiff}
for large $T$
corresponds to the highest eigenvalue of this deformed generator
with its right positive eigenvector $ {\tilde r}^{\pm}_k( x) $
\begin{eqnarray}
  \mu(k) {\tilde r}^+_k( x)  &&   =  
        - \frac{d}{dx}   \left[ v_+(x) {\tilde r}^+_k( x )  \right]  
+  \left[ k \alpha_+ ( x )+ k \omega_+(x) v_+(x) -\gamma_+(x)  \right]   {\tilde r}^+_k( x)
 + \gamma_- (x) e^{k \beta_-(x) }{\tilde r}^-_k( x)
 \nonumber \\
  \mu(k) {\tilde r}^-_k( x) &&   =  
   \frac{d}{dx}   \left[ v_-(x) {\tilde r}^-_k( x )  \right]
    + \gamma_+(x) e^{k \beta_+(x) } {\tilde r}^+_k( x)
+  \left[ k \alpha_- ( x ) -  k \omega_-(x) v_-(x)-\gamma_-(x)  \right]   {\tilde r}^-_k( x)
  \label{tright}
\end{eqnarray}
and its left positive eigenvector $ {\tilde l}^{\pm}_k( x) $
\begin{eqnarray}
  \mu(k) {\tilde l}^+_k( x)  &&   =      
v_+(x)  \frac{d}{dx}   {\tilde l}^+_k( x) 
+  \left[k \alpha_+ ( x ) + k \omega_+(x) v_+(x) - \gamma_+(x) \right]   {\tilde l}^+_k( x)
 + \gamma_+(x) e^{k \beta_+(x) } {\tilde l}^-_k( x)
 \nonumber \\
  \mu(k) {\tilde l}^-_k( x)  &&   =      
- v_-(x)     \frac{d}{dx}  {\tilde l}^-_k( x) 
+ \gamma_-(x) e^{k \beta_-(x) } {\tilde l}^+_k( x)
+  \left[ k \alpha_- ( x )-  k \omega_-(x) v_-(x)- \gamma_-(x)  \right]   {\tilde l}^-_k( x) 
  \label{tleft}
\end{eqnarray}
with the normalization
\begin{eqnarray}
\int_{-\infty}^{+\infty}d  x \left[  {\tilde l}^+_k( x) {\tilde r}^+_k( x) +  {\tilde l}^-_k( x) {\tilde r}^-_k( x)\right] =1
  \label{tnorma}
\end{eqnarray}

\subsection { The corresponding conditioned process obtained via the generalization of Doob's h-transform }

The Doob conditioned process is conserved and very similar to the initial dynamics of Eq. \ref{run}
\begin{eqnarray}
 \frac{ \partial {\tilde{\tilde P}}_+( x,t)   }{\partial t}  &&   =  
 -     \partial_x   \left[ v_+(x) {\tilde{\tilde P}}_+( x ,t)  \right]  
 -{\tilde{\tilde \gamma}}_k^+(x)    {\tilde{\tilde P}}_+( x,t)
 + {\tilde{\tilde \gamma}}_k^-(x)  {\tilde{\tilde P}}_-( x,t)
 \nonumber \\
  \frac{ \partial {\tilde{\tilde P}}_-( x,t)   }{\partial t}  &&   =  
        \partial_x  \left[ v_-(x) {\tilde{\tilde P}}_-( x,t )  \right]
    + {\tilde{\tilde \gamma}}_k^+(x)  {\tilde{\tilde P}}_+( x,t)
 -{\tilde{\tilde \gamma}}_k^-(x)    {\tilde{\tilde P}}_-( x,t)
\label{doob}
\end{eqnarray}
since the only difference is that
the two effective switching rates depend on the two functions $\beta_{\pm}(x) $ and
on the left eigenvector ${\tilde l}^{\pm}_k( x) $ of Eq. \ref{tleft}
\begin{eqnarray}
{\tilde{\tilde \gamma}}_k^+(x)  && \equiv  \gamma_+(x) e^{k \beta_+(x) }  \frac{{\tilde l}^-_k( x)}{{\tilde l}^+_k( x)}
\nonumber \\
{\tilde{\tilde \gamma}}_k^-(x)  && \equiv  \gamma_-(x) e^{k \beta_-(x) }  \frac{{\tilde l}^+_k( x)}{{\tilde l}^-_k( x)}
  \label{doobgammaeff}
\end{eqnarray}
The steady state of the conditioned process of Eq. \ref{doob}
involve the left and the right eigenvectors of Eq. \ref{tright} and Eq. \ref{tleft}
\begin{eqnarray}
{\tilde{\tilde \rho}}^{+*}_k( x)   &&   =  {\tilde l}^+_k( x){\tilde r}^+_k( x)
\nonumber \\
{\tilde{\tilde \rho}}^{-*}_k( x)   &&   =  {\tilde l}^-_k( x){\tilde r}^-_k( x)
\label{doobrho}
\end{eqnarray}
with the corresponding steady spatial currents 
\begin{eqnarray}
{\tilde{\tilde j}}^{+*}_k( x)   &&   = v_+(x) {\tilde{\tilde \rho}}^{+*}_k( x) = v_+(x) {\tilde l}^+_k( x){\tilde r}^+_k( x)
\nonumber \\
{\tilde{\tilde j}}^{-*}_k( x)   &&   = - v_-(x) {\tilde{\tilde \rho}}^{-*}_k( x)     = - v_-(x){\tilde l}^-_k( x){\tilde r}^-_k( x)=
- {\tilde{\tilde j}}^{+*}_k( x)
\label{doobj}
\end{eqnarray}
and the corresponding steady switching flows
\begin{eqnarray}
{\tilde{\tilde Q}}^{+*}_k( x)   &&   = {\tilde{\tilde \gamma}}_k^+(x)  {\tilde{\tilde \rho}}^{+*}_k( x)  
= \frac{{\tilde{\tilde \gamma}}_k^+(x)}{ v_+(x)}{\tilde{\tilde j}}^{+*}_k( x)
\nonumber \\
{\tilde{\tilde Q}}^{-*}_k( x)   &&   = {\tilde{\tilde \gamma}}_k^-(x)  {\tilde{\tilde \rho}}^{-*}_k( x)  
= \frac{{\tilde{\tilde \gamma}}_k^-(x)}{ v_-(x)}{\tilde{\tilde j}}^{+*}_k( x)
\label{doobq}
\end{eqnarray}

\subsection { Equivalence with the optimization from the Level 2.5 }

The consistency between the approach based on the optimization of the Level 2.5 (see section \ref{sec_additive})
and the alternative approach based on the deformed generator with the corresponding Doob conditioned process
yields that
the steady state observables of the conditioned Doob process described above
should coincide with the optimal solution found via the contraction of the Level 2.5 in section \ref{sec_additive}.
In the present case, this means that
the optimal current $ j^{+opt}_{k}(x) $ and the optimal switching flows $Q^{\pm opt}_{k}(x) $ 
of section \ref{sec_additive}
correspond to the Doob steady current of Eq. \ref{doobj}
and to the Doob steady switching flows of Eqs \ref{doobq}
\begin{eqnarray}
j^{+opt}_{k}(x) && = {\tilde{\tilde j}}^{+*}_k( x)   
\nonumber \\
Q^{\pm opt}_{k}(x) && = {\tilde{\tilde Q}}^{\pm *}_k( x)  
\label{doobcorresjq}
\end{eqnarray}
The identification between the Doob switching rates of Eq. \ref{doobgammaeff}
and the effective switching rates of the optimal solution of Eq. \ref{qoptstep3}
yields that the Lagrange multiplier $\lambda_k(x)$
corresponds to the logarithm of the ratio of the two components ${\tilde l}^{\pm}_k( x) $ of the left eigenvector of Eq. \ref{tleft}
\begin{eqnarray}
\lambda_k(x)  = \ln \left( \frac{{\tilde l}^+_k( x)}{{\tilde l}^-_k( x)} \right)
  \label{lambdaratio}
\end{eqnarray}
Using Eq. \ref{tleft}, one obtains that the ratio of Eq. \ref{lambdaratio}
 satisfies the non-linear differential equation
\begin{small}
\begin{eqnarray}
&& \frac{d\lambda_k(x)}{dx}    = 
\frac{\frac{d{\tilde l}^+_k( x)}{dx}}{{\tilde l}^+_k( x)} - \frac{\frac{d{\tilde l}^-_k( x)}{dx}}{{\tilde l}^-_k( x)}
 \label{lambdaratioderi}
\\
&& = -  \frac{\gamma_+(x) e^{k \beta_+(x) } }{v_+(x)} e^{- \lambda_k(x)} 
 -  \frac{\gamma_-(x) e^{k \beta_-(x) } }{v_-(x)}   e^{ \lambda_k(x)}
+ \frac{\gamma_+(x) +  \mu(k)  }{v_+(x)} 
+ \frac{  \gamma_-(x) + \mu(k)   }{v_-(x)} 
- k \left[ \omega_+(x) -  \omega_-(x) + \frac{\alpha_{+}( x)}{v_+(x) } + \frac{\alpha_{-}( x)}{v_-(x) } \right]
\nonumber  
\end{eqnarray}
\end{small}
that indeed coincides with Eq. \ref{lagrangederi3expli} in terms of the notation $\omega(x)$ introduced in Eq. \ref{omega}.


\section{ Link between the Level 2.75 and the Level 2.5 }

\label{app_contraction2.75to2.5}

In this Appendix related to section \ref{sec_intervals}, we explain 
the link between the Level 2.75 and the Level 2.5.
The comparison between the Level 2.5 of Eq. \ref{ld2.5rhoq} 
and the Level 2.75 of Eq. \ref{ld2.75}
yields that the
conditional probability to see the empirical densities of intervals $q_{\pm} (.,.)$
 once the other empirical observables $[j_+(.),  Q_{\pm}(.) ]$ are given reads
\begin{eqnarray}
P_T^{conditional} [ q_{\pm} ( . , .) \vert j_+(.),  Q_{\pm}(.) ]  
&& \equiv \frac{  P_T[   j_+(.), Q_{\pm}(.), q_{\pm} (.,.)   ]  }
{  P_T[  j_+(.) , Q_{\pm}(.) ]   } \opsimeq_{T \to +\infty}  e^{ \displaystyle -T   I^{conditional} [ j_+(.), Q_{\pm}(.), q_{\pm} (.,.) ]}
\label{ld2.75condi}
 \\
&&  
\left[ \prod_{ x_R } \delta \left(  \int_{-\infty}^{x_R} dx_L q_{+} (x_R,x_L)  - Q_{+}( x_R)  \right)  \right]
\left[ \prod_{ x_R } \delta \left( \int_{-\infty}^{x_R} dx_L q_{-} (x_L,x_R) - Q_{+}( x_R)   \right)  \right]
\nonumber \\
&&\left[ \prod_{ x_L } \delta \left(   \int_{x_L}^{+\infty} dx_R q_{+} (x_R,x_L) - Q_{-}( x_L)   \right)  \right]
\left[ \prod_{ x_L } \delta \left(   \int_{x_L}^{+\infty} dx_R q_{-} (x_L,x_R) -   Q_{-}( x_L)  \right)  \right]
\nonumber
\end{eqnarray}
The conditional rate function reads
\begin{small}
\begin{eqnarray}
&&  I^{conditional} [  j_+(.), Q_{\pm}(.), q_{\pm} (.,.)] 
   =
  I_{2.75} [  j_+(.), Q_{\pm}(.), q_{\pm} (.,.)] -  I_{2.5}[ j_+,  Q_{\pm}(.) ]  
 \label{rate2.75conditional}
 \\
  && =     \int_{-\infty}^{+\infty} dx_L \int_{x_L}^{+\infty} dx_R
\left[ q_{+} (x_R,x_L)  \ln \left( \frac{ q_{+} (x_R,x_L) }{ \frac{\gamma_+(x_R) }{v_+(x_R)} \ e^{ - \int_{x_L}^{x_R} dx \frac{\gamma_+( x )}{v_+(x)}  } Q_-(x_L)  }   \right)
+ q_{-} (x_L,x_R)   \ln \left( \frac{ q_{-} (x_L,x_R)  }{ \frac{\gamma_-(x_L) }{v_-(x_L)} \ e^{ - \int_{x_L}^{x_R} dx \frac{\gamma_-( x )}{v_-(x)}  } Q_+(x_R) }   \right)  
 \right]
  \nonumber \\
  && 
    -   \int_{-\infty}^{+\infty}d  x
\left[  Q_{+}( x)    \ln \left( \frac{  Q_{+}( x)   }{     \frac{\gamma_+(x)}{v_+(x) }  j_+(x) }  \right) 
 -   Q_{+}( x)   +    \frac{\gamma_+(x)}{v_+(x) }  j_+(x)    \right]  
 -  \int_{-\infty}^{+\infty}d  x
\left[  Q_{-}( x)   \ln \left( \frac{  Q_{-}( x)   }{    \frac{\gamma_-(x)}{v_-(x) }  j_+(x) }  \right) 
 -   Q_{-}( x)   +   \frac{\gamma_-(x)}{v_-(x) }  j_+(x)  \right]   
    \nonumber
\end{eqnarray}
\end{small}

The explicit form of the kernel $W_{\pm}(  .  \vert .  ) $ of Eqs \ref{emoins} and \ref{eplus},
can be translated for
 the effective kernels $ \hat W_{\pm}(  .  \vert .  ) $ associated to the effective switching rates 
$\hat \gamma_{\pm}(x)  = v_{\pm}(x) \frac{  Q_{\pm}(x)  }{  j_+(x) }$
 of Eq. \ref{hatgamma}
\begin{eqnarray}
\hat W_-(x_L \vert x_R) 
&& = \frac{ \hat\gamma_-(x_L) }{v_-(x_L)} \ e^{\displaystyle - \int_{x_L}^{x_R} dx \frac{\hat\gamma_-( x )}{v_-(x)}  }
=
\frac{  Q_{-}(x_L)  }{  j_+(x_L) } \ e^{\displaystyle - \int_{x_L}^{x_R} dx \frac{  Q_{-}(x)  }{  j_+(x) }  }
\nonumber \\
\hat W_+(x_R \vert x_L) 
&& = \frac{\hat\gamma_+(x_R) }{v_+(x_R)} \ e^{\displaystyle - \int_{x_L}^{x_R} dx \frac{\hat \gamma_+( x )}{v_+(x)}  }
= \frac{  Q_{+}(x_R)  }{  j_+(x_R) } \ e^{\displaystyle - \int_{x_L}^{x_R} dx \frac{  Q_{+}(x)  }{  j_+(x) }  }
\label{hatWkernel}
\end{eqnarray}
These effective kernels $ \hat W_{\pm}(  .  \vert .  ) $ are useful together with the constraints
to rewrite the 
conditional rate function of Eq. \ref{rate2.75conditional} as
\begin{small}
\begin{eqnarray}
&&  I^{conditional} [  j_+(.), Q_{\pm}(.), q_{\pm} (.,.)]  
\label{rateiconditionalhat}
 \\
&& = \int_{-\infty}^{+\infty} dx_L \int_{x_L}^{+\infty} dx_R
\left[ q_{+} (x_R,x_L)  \ln \left( \frac{ q_{+} (x_R,x_L) }{ \hat W_+(x_R \vert x_L) Q_-(x_L)  }   \right)
+ q_{-} (x_L,x_R)   \ln \left( \frac{ q_{-} (x_L,x_R)  }{ \hat W_-(x_L \vert x_R) Q_+(x_R) }   \right)  
 \right]
 \nonumber \\
&& = \int_{-\infty}^{+\infty} dx_L \int_{x_L}^{+\infty} dx_R
\left[ q_{+} (x_R,x_L)  \ln \left( \frac{ q_{+} (x_R,x_L) }{ \frac{  Q_{+}(x_R)  }{  j_+(x_R) } \ e^{ - \int_{x_L}^{x_R} dx \frac{  Q_{+}(x)  }{  j_+(x) }  } Q_-(x_L)  }   \right)
+ q_{-} (x_L,x_R)   \ln \left( \frac{ q_{-} (x_L,x_R)  }{ \frac{  Q_{-}(x_L)  }{  j_+(x_L) } \ e^{ - \int_{x_L}^{x_R} dx \frac{  Q_{-}(x)  }{  j_+(x) }  } Q_+(x_R) }   \right)  
 \right]  
\nonumber
\end{eqnarray}
\end{small}
This factorized form shows that this conditional rate function vanishes for the optimal values
of the empirical density of intervals
\begin{eqnarray}
q_{+}^{opt} (x_R,x_L) && =\frac{  Q_{+}(x_R)  }{ j_+(x_R) } \ e^{\displaystyle - \int_{x_L}^{x_R} dx \frac{  Q_{+}(x)  }{  j_+(x) }  } Q_-(x_L) 
\nonumber \\
q_{-}^{opt} (x_L,x_R) && = \frac{  Q_{-}(x_L)  }{ j_+(x_L) } \ e^{ \displaystyle - \int_{x_L}^{x_R} dx \frac{  Q_{-}(x)  }{  j_+(x) }  } Q_+(x_R)
\label{qspoptimal}
\end{eqnarray}
once the other one-position empirical observables $[  j_+(.), Q_{\pm}(.)] $ are given.
Taking into account the stationarity constraint of Eq. \ref{jumpstatiofinal}
to replace $ Q_{+}( x)  =   - \frac{d j_+( x)  }{dx}  + Q_{-}( x)$ in the exponential of the first Eq \ref{qspoptimal}
\begin{eqnarray}
q_{+}^{opt} (x_R,x_L) && =\frac{  Q_{+}(x_R) Q_-(x_L) }{ j_+(x_R) } \ 
e^{\displaystyle + \int_{x_L}^{x_R} dx   \frac{d \ln j_+( x)  }{dx} - \int_{x_L}^{x_R} dx \frac{ Q_{-}( x)  }{  j_+(x) } 
   }  
   \nonumber \\
   && =\frac{  Q_{+}(x_R) Q_-(x_L) }{ j_+(x_L) } \ 
e^{\displaystyle  - \int_{x_L}^{x_R} dx \frac{ Q_{-}( x)  }{  j_+(x) } 
   } = q_{-}^{opt} (x_L,x_R)
\label{qspoptimalequiv}
\end{eqnarray}
one obtains that the two optimal values of Eq. \ref{qspoptimal} actually coincide.
Since they satisfy the constraints of Eq. \ref{ld2.75condi}, they correspond to the optimal solutions
for the contraction of the Level 2.75 to recover the Level 2.5.


\end{document}